\documentclass[10pt,onecolumn,letterpaper]{article}

\usepackage{times}
\usepackage{epsfig}
\usepackage{graphicx}
\usepackage{amsmath}
\usepackage{amssymb}
\usepackage{array}
\usepackage{adjustbox}
\usepackage{array}
\usepackage{cellspace}
\usepackage{hhline}

\graphicspath{{figuras/}}
\usepackage[utf8]{inputenc} 
\usepackage{multirow}
\usepackage{placeins}
\usepackage{lineno}
\usepackage{color}
\usepackage{hyperref}
\usepackage{booktabs}
\usepackage{longtable} 
\usepackage{url}
\usepackage{subfig}   
\pdfoutput=1

\newcommand{\method}[1]{\texttt{#1}}
\usepackage[table]{xcolor}

\begin{document}
	\title{The OCR-PT-CT Project: Semi-Automatic Recognition of Ancient Egyptian Hieroglyphs Based on Metric Learning}
	
	\author{
		David Fuentes-Jimenez$^{1}$,
		Daniel Pizarro$^{1}$,
		Álvaro Hernández$^{1}$,\\
		Adin Bartoli$^{1}$,
		Laura de Diego-Otón$^{1}$,
		Sira Palazuelos-Cagigas$^{1}$,\\
		César Guerra Méndez$^{2}$,
		Carlos Gracia Zamacona$^{2}$\\[1ex]
		$^{1}$Electronics Department, University of Alcalá\\
		$^{2}$Department of History and Philosophy, University of Alcalá\\
		Alcalá de Henares, Madrid, Spain\\
		{\tt\small d.fuentes@uah.es, carlos.gracia@uah.es}
	}
	
	\maketitle
\begin{abstract}
Digital humanities are significantly changing the way that many Egyptologists deal with ancient Egyptian texts. The design and integration of supporting computer tools have implied a faster and different approach to how to treat, exploit and disseminate ancient Egyptian texts. In this context, the OCR-PT-CT project proposes a recognition method of ancient Egyptian hieroglyphs, based on imported images of the Coffin Texts (CT) available from the synoptic edition by Adriaan de Buck (1935-1961) and the edition of Pyramid Texts (PT) copies on Middle Kingdom coffins prepared by James Allen (2006). The proposal is capable of identifying the existing hieroglyphs, and transcribing them into the corresponding Gardiner’s codes. Furthermore, a web tool allows to organize them according to spells and witnesses, and store all the resulting information into a csv file ready to be integrated with the MORTEXVAR project dataset. This dataset is an ongoing project gathering the Coffin Texts from all Middle Kingdom sources, together with the associated metadata and the corresponding transliteration and translation, as a critical basis for further research on the topic. The recognition of hieroglyphs has been approached from two different points of view. On one hand, a well-known neural network, Mobilenet, has been trained to identify up to $140$ different hieroglyphs (classes), achieving a global accuracy of $93.87$\% roughly, showing a significant degradation of the performance when dealing with hieroglyphs that do not have enough samples for training (unbalanced dataset). On the other hand, a novel approach based on metric learning has been designed, which provides the flexibility to deal with new signs not seen during training in a flexible way, or with those classes with a reduced number of samples (data-limited scenarios). This alternative achieves a global accuracy of $97.70$\%, and allows a larger number of hieroglyphs to be recognised. Among the evaluated alternatives, the final system adopts the proposed Deep Metric Learning approach as the default classifier due to its superior performance under class imbalance and its flexibility to incorporate new hieroglyphs.
\end{abstract}

\section{Introduction}

Optical Character Recognition (OCR)  has become a common approach when dealing with printed or handwritten documents that should be converted into readable formats by computation systems. OCR techniques have been developed from some decades ago \cite{Mori92}, varying from alternatives focused on single characters to others considering the whole text as a single input. Overall, the different OCR proposals already available perform suitably with modern languages \cite{Memon20}, where huge amounts of data can be complied for training. Nevertheless, some issues arise when working with ancient languages (e.g., Ancient Egyptian): the corpora are often reduced, the conservation of these documents often implies additional challenges, and the complexity of the handwriting involved cannot be compared to modern Western languages and alphabets. 

Since the emergence of deep computational architectures, different works have been and are currently being developed in the field \cite{sproat2010}. The relevance of how a computational approach can enhance some tasks in the study of an ancient language, such as linguistic analysis, translation or decipherment, is already mentioned in \cite{sommerschield2023}, where the importance of interdisciplinary teams with historians, linguistics and data scientists, is the key to successfully apply machine learning in the domain. For instances, in \cite{braovic2024} a study about decipherment in Bronze Age Aegean and Cypriot Scripts is carried out, highlighting the lack of uniform and standardised datasets, which are often too reduced to allow the application of deep learning techniques. However, it also remarks how deep learning may perform better at seeking for links between languages and scripts from a chronological and geographical point of view. Additionally, open challenges such as the development of robust post-OCR processing techniques have been driven by recent developments in machine learning, especially deep learning \cite{Nguyen21}. Accordingly, artificial intelligence (AI) systems have allowed in last years the appearance of recent implementations for ancient Egyptian texts \cite{rosmorduc2015, zamacona2021, lucarelli2023, fuentes2025, KE2025, WANG2021}.

In light of its increasing relevance, numerous research initiatives have been developed in recent years. For example, the project Book of the Dead in 3D \cite{lucarelli2023b} outstands in the complementary use of photogrammetry for the 3D reconstruction of the supports and text encoding and annotation, so to cover their transcription, transliteration and translation. This procedure conveniently combines museum heritage conservation with text edition and dissemination \cite{lucarelli2019}. Additionally, the project Altägyptische Kursivschriften \cite{konrad2023}, combines philological research and digital humanities methods by applying OCR to the hieratic script (a shorthand that evolved from ancient Egyptian hieroglyphs). It was one of the most innovative in the field, since it used a convolutional neural network to classify the hieratic characters. The Coptic OCR project \cite{miyagawa2019} has shown a very high reliability of the OCR process with Coptic manuscripts: between $98.5$~\% and $99.6$~\% (depending on manuscript’s handwritings). These texts are written in an alphabetic script very similar to Greek to convey the last phase of the Egyptian language (since the third century CE, and still alive as the liturgical language of the Christian church in Egypt). This project also applies artificial intelligence techniques to learn a language model and launch predictions of characters according to context. Furthermore, the Hieroglyphics Initiative project consists in applying OCR techniques on a collection of facsimiles kept in the Berlin-Brandenburg Academy of Sciences \cite{Ubisoft17}. It is developed by Psycle \cite{Psycle17} with the tool Fabricius, hosted and disseminated by Google Arts \& Experiments \cite{Google17, kelly2021}. Finally, similar approaches to Egyptian hieroglyphic OCR include Franken and Gemert \cite{franken2013}, Barucci et al. \cite{barucci2021}, Elnabawy et al. \cite{elnabawy2018} \cite{elnabawy2021}, and S. Rosmorduc \cite{rosmorduc2020}. Recent work has specifically targeted the segmentation stage with Mask R-CNNs \cite{Guidi2023Segmentation}, and has even combined CNN segmentation with hyperspectral imaging to improve readability on degraded artefacts \cite{Cucci2024HSI}. Beyond Egyptology, a relevant project is Hu’s to Mayan glyphs, which develops an intelligent system based on computer vision \cite{hu2017}.

Complementarily to these OCR-based approaches and research initiatives, other encoding-annotation approaches are of a special relevance. For instances, the project Machine Readable Texts for Egyptologists \cite{Jauhiainen21} studies the ways of publishing ancient Egyptian texts in machine-readable form and aims at producing encoded hieroglyphic texts from the New Kingdom. The project Strukturen und Transformationen des Wortschatzes der ägyptischen Sprache: Text- und Wissenskultur im Alten Ägypten \cite{topmann21} is using the more comprehensive dictionary of ancient Egyptian in electronic format (Thesaurus Linguae Aegyptiae) to annotate texts. In the same research line, Prof. Mark-Jan Nederhof’s project PhilologEg \cite{Nederhof02}, adopts a Unicode-based approach for Egyptian texts encoding and further approaching of the ancient Egyptian writing system through a probabilistic model \cite{nederhof2017}. Finally, two major projects use an encoding standard code of transcription and transliteration known as Manuel de Codage (MdC) \cite{Buurman88, Grimal00}, that is based on the Gardiner’s sign list codes. These projects are in the process to provide annotated text corpora of later date: the Ramses online project \cite{polis2015, winand2015}, of Late Egyptian texts (ca. 1500-700 BC); and the Demotic Palaeographical Database Project, of Demotic texts (ca. 700 BC – 500 CE) \cite{Quack23}.

In this context, the OCR-PT-CT project \cite{Gracia22a} proposes the semi-automatic transcription of ancient Egyptian hieroglyphs from the Coffin Texts (CT) and the Pyramid Texts (PT). Consequently, the OCR-PT-CT project proposes a tool that is capable of processing images with texts in order to identify the hieroglyphs, transcript them to Gardiner’s codes \cite{Buurman88}, and store them for further processing, according to source, support, witness and spell. Specifically, the main novelties of this work are:

\begin{itemize}
	\item The definition of a Deep Metric Learning model, capable of identifying Egyptian hieroglyphs from a digital source \cite{Buck61}, even under strong class imbalance.
	\item The detailed comparison of the proposal with other classic approaches, such as a Traditional Machine Learning baseline and an End-to-End Deep CNN Classifier.
	\item The successful experimental validation of the proposal by training the system with the digital source \cite{Buck61} and assessing it with some ground-truth pages, unseen by the system previously.
\end{itemize}

The detailed comparison with classic approaches (Traditional ML and an end-to-end CNN) is included as a benchmark; however, the proposed system adopts Deep-MML as the default recognition backend. All data and code developed in this work will be released for public use. The datasets will be made available through the MORTEXVAR project website, while the source code will be shared via GitHub and the Kaggle data science community. The outcomes of the entire process are regularly reviewed by expert Egyptologists and stored into a csv file ready to be integrated with the MORTEXVAR project dataset (TM-CT project, \url{https://www.mortexvar.com/tm-ct}). Consequently, the transcription of the hieroglyphs, which is provided by the OCR-PT-CT project, will be available to be linked with the transliteration (i.e. the alphabetic representation of the ancient Egyptian words’ consonantal structure) and the translation, which is provided by the MORTEXVAR project \cite{gracia2013, zamacona2020}. 

The rest of the manuscript is organized as follows: Section \ref{sec2} provides a general description of the proposal; Section \ref{sec3} describes the methods dedicated to character recognition; Section \ref{sec4} defines how the character classification is carried out; Section \ref{sec5} provides some experimental results; and, finally, conclusions are discussed in Section \ref{sec7}.

\section{General Overview of the Proposal}
\label{sec2}

\begin{figure}
	\centering
	\includegraphics[width=1.0\linewidth]{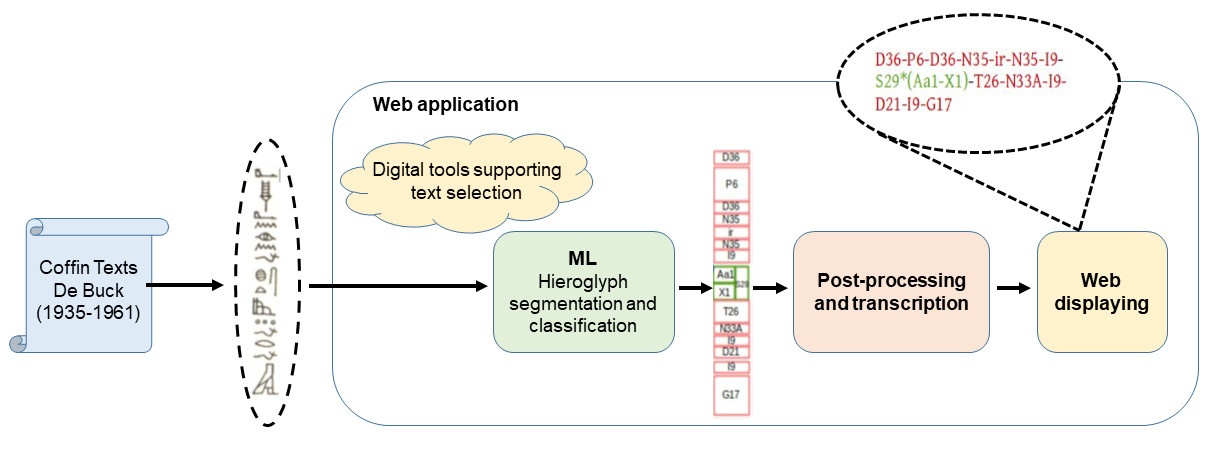} 
	\caption{General block diagram of the proposed hieroglyph segmentation and classification, starting from a text available in a digital edition until obtaining the text transcription  in the MdC nodes.}\label{fig_1}
\end{figure}

This work proposes an intelligent system for the semi-automatic transcription of hieroglyphic characters from ancient Egyptian texts. The proposal focuses initially on hieroglyphs found in digital editions of the Coffin Texts (CT) available in \cite{Buck61} and copies of the Pyramid Texts (PT) in Middle Kingdom documents \cite{Allen06}. The CT are the largest corpus of funerary texts ever produced in ancient Egypt, and they were written during the Middle Kingdom (ca. 2000-1500 BCE) \cite{willems2014, gracia2024}. The OCR-PT-CT also uses the corpus of Middle Kingdom copies of PT, an earlier funerary corpus (ca. 2500-2000 BCE). These two corpora are closely related to each other in the topics covered, as well as in the vocabulary, grammar and specific word-spelling that were employed. The cultural, linguistic and graphemic unity of the PT-CT is one of the strongest assets of the project from the Egyptological point of view. Fig.~\ref{fig_1} depicts a general block diagram of the approach. 
The entire process has been integrated into a web application designed to provide digital tools for managing the aforementioned digital editions and to allow users to select specific image regions containing hieroglyphs of interest. The structure has been developed around a main block responsible for segmentation of hieroglyphs in images captured from the aforementioned digital editions and character classification, whose results are subsequently used for post-processing and transcription.

Building on the general overview presented above, the following describes the complete procedure in detail. After region selection, a dedicated \textbf{segmentation module} processes the image to isolate individual hieroglyphs. This involves several steps: binarizing the image using adaptive thresholding, identifying potential symbol regions via connected components analysis, filtering these components based on size to remove noise and irrelevant elements, and optionally grouping the remaining components into vertical columns to reflect the text layout. Following segmentation, the isolated hieroglyph candidates are passed to the \textbf{classification stage}. Our primary proposed method for identifying the corresponding code from the MdC is based on deep metric learning. This involves using a Convolutional Neural Network (CNN) trained to generate unique numerical fingerprints for each symbol, enabling classification by comparing an unknown symbol's fingerprint to the average fingerprint of known MdC codes. For comprehensive evaluation, the system framework also supports comparative classification methods, including traditional machine learning approaches using predefined visual features and direct end-to-end classification using a fine-tuned CNN. Unless otherwise stated, the web/desktop application runs Deep Metric Learning classifier as the default classifier, other approximations are provided for comparison and ablation purposes. Finally, the last stage is dedicated to \textbf{post-processing}, potential refinement of the transcription results, and their visualization within a dedicated web application interface developed for that purpose. 

\subsection{Corpus Description and Processing}
\label{sec2_1}

\begin{figure}
	\centering
	\includegraphics[width=1.0\linewidth]{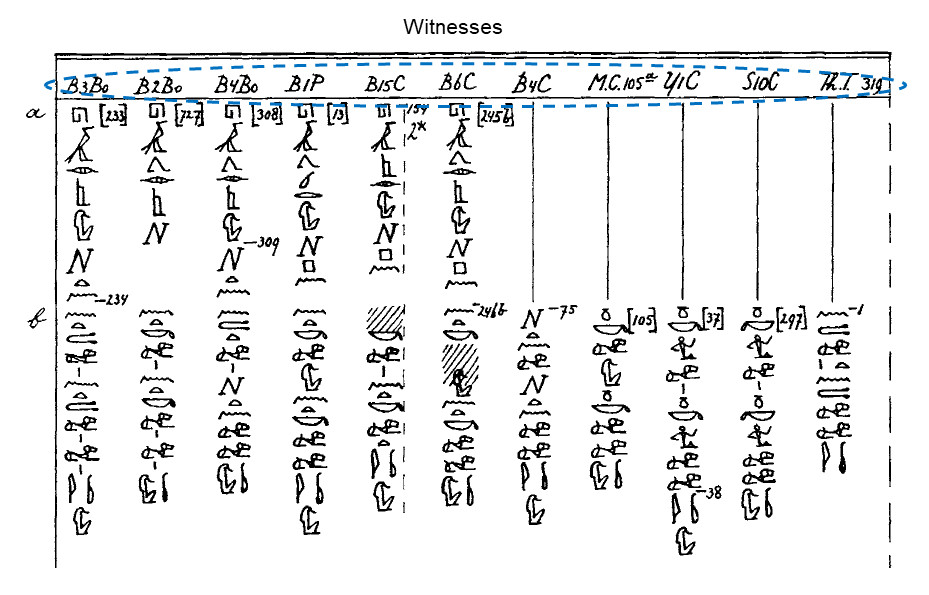} 
	\caption{General view of the data corpus used in this work, particularly the spell no. 1 in \cite{Buck61}.}\label{fig_2}
\end{figure}

The data corpus involved hereinafter in the training of the proposed hieroglyph segmentation and classification methods is based on the book The Egyptian Coffin Texts, by Adriaan De Buck (The University of Chicago Press) \cite{Buck61}. This compiles a set $163$ spells, organised into two PDF volumes. Every spell is arranged in parallel vertical columns, where each column corresponds to a different witness (or coffin). In this way, parallel columns tend to convey the same information, frequently using exactly the same hieroglyphs. As an example, Fig. \ref{fig_2} provides a general view of spell no. 1 in \cite{Buck61}, where it is possible to observe until eleven different witnesses (note that the codes surrounded by the blue ellipse in Fig. \ref{fig_2} are the notations used to refer to the coffins/witnesses).

\begin{figure}[!b]
	\centering
	
	\subfloat[A1]{%
		\includegraphics[width=0.10\linewidth]{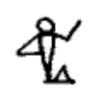}
		\includegraphics[width=0.10\linewidth]{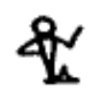}
		\includegraphics[width=0.10\linewidth]{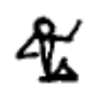}
	}
	\hspace{0.02\linewidth}
	\subfloat[Y1]{%
		\includegraphics[width=0.10\linewidth]{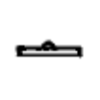}
		\includegraphics[width=0.10\linewidth]{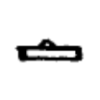}
		\includegraphics[width=0.10\linewidth]{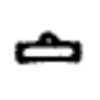}
	}
	
	\vspace{2mm}
	
	\subfloat[G1]{%
		\includegraphics[width=0.10\linewidth]{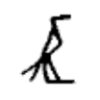}
		\includegraphics[width=0.10\linewidth]{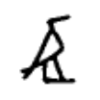}
		\includegraphics[width=0.10\linewidth]{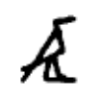}
	}
	\hspace{0.02\linewidth}
	\subfloat[M17]{%
		\includegraphics[width=0.10\linewidth]{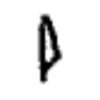}
		\includegraphics[width=0.10\linewidth]{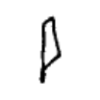}
		\includegraphics[width=0.10\linewidth]{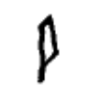}
	}
	
	\caption{Three samples of the hieroglyphs with the specified MdC code available in the dataset generated from \cite{Buck61}.}
	\label{fig_3}
\end{figure}

Every column was selected manually by means of an ad-hoc web tool developed to alleviate the workload to specialists. These captured images become the input of the character segmentation method described in Section~\ref{sec3}. Afterwards, the resulting isolated hieroglyphs are curated by the Egyptologists in the research group, so they are correctly labelled to form a suitable dataset that can be used to train the character classification methods proposed in this work. This final dataset is composed by $262$ different categories, corresponding to the codes or classes defined in the MdC by Gardiner. It is worth noting that the number of samples per class is quite variable, since some hieroglyphs are very frequent (more than a thousand samples are available in these cases), whereas other symbols are scarce. Furthermore, when considering only hieroglyph classes with more than 140 samples available for training, validation, and testing, the number of distinct classes is reduced to a total of 140 hieroglyphs, which significantly complicates the training process. Finally, every sample is a PNG image file, with a size of $100\times 100$~pixels and a depth of $24$~bits. Fig.~\ref{fig_3} shows examples of three different hieroglyphs, manually labelled and associated with the MdC categories A1, Y1, G1 and M17, respectively.

A last aspect to take into consideration is how data are grouped later during the training of the proposed intelligent systems. Note that the available classes/hieroglyphs will be divided into three subsets, corresponding to the phases of training, validation and tests. Nevertheless, in order to to discard any data leakage, the goal is to avoid the proposed systems to see any hieroglyphs in the test phase coming from the same pages as other signs involved in training. For that purpose, ten pages from the original digital edition \cite{Buck61} have been extracted and reserved to be employed in the three phases previously mentioned, so a further ground-truth test can be carried out with the trained systems.

\begin{table}[]
	\caption{MORTEXVAR dataset splits and samples.}
	\centering
	\begin{tabular}{|c|c|c|}
		\hline
		\textbf{Modality}             & \textbf{Samples} & \textbf{\%} \\ \hline
		\textbf{Train}                & 26653            & 61.95\%     \\ \hline
		\textbf{Validation}           & 5874             & 13.65\%     \\ \hline
		\textbf{Test(15\%)}           & 5874             & 13.65\%     \\ \hline
		\textbf{Test(isolated pages)} & 4622             & 10.74\%     \\ \hline
	\end{tabular}
	\label{tab:dataset}
\end{table}

\subsection{Web and Local application}

To make the proposed OCR-PT-CT pipeline usable by domain experts, we provide two complementary application modalities that expose the same core functionality: (i) a web-based interface implemented with \textit{Streamlit} \cite{streamlit2025}, and (ii) a standalone desktop graphical user interface (GUI) implemented with \textit{Tkinter} \cite{lundh1999introduction}. Both modalities integrate (a) facsimile browsing, (b) region-of-interest (ROI) selection, (c) execution of the hieroglyph segmentation and classification modules, and (d) structured export of the resulting transcription and metadata. Importantly, the full workflow runs on CPU, enabling execution on standard computers without dedicated GPU hardware.

\begin{figure}
	\centering
	\includegraphics[width=1.0\linewidth]{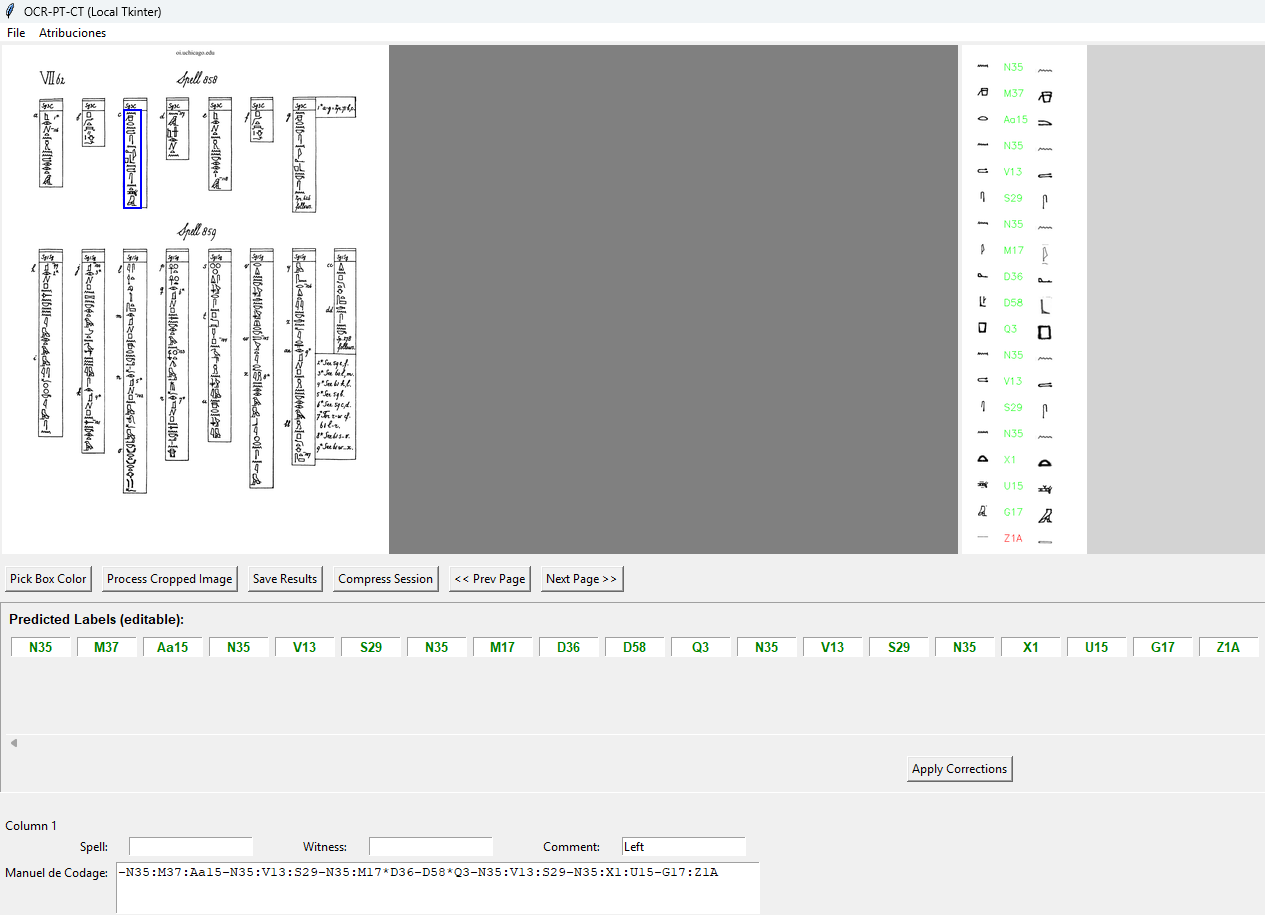} 
	\caption{Example view of the OCR local application, loading an image, selecting a ROI, and transcribing it.}\label{fig_10}
\end{figure}

In both modalities, the typical workflow consists of the following steps:
(1) the user selects a corpus source (e.g., spell and witness/page) and loads the corresponding facsimile;
(2) the user draws one or more ROIs covering a vertical column (or any target block) of hieroglyphs;
(3) the segmentation module extracts individual sign candidates;
(4) each candidate is classified into a Gardiner/MdC code using the selected classifier (Deep-MML by default);
(5) the interface presents the ordered list of predicted signs together with auxiliary information (e.g., similarity/confidence scores and/or bounding boxes) to support rapid expert review; and
(6) the output is exported as a CSV file encoding spell/witness/token information and predicted code(s), ready to be ingested by the MORTEXVAR data pipeline. We show in Figure \ref{fig_10} an example of the local application created using Tkinter libraries.

Both applications share the same Python processing backend (segmentation, feature extraction, and classification), ensuring consistent outputs across interfaces.
The Streamlit modality provides a browser-based experience that can be launched locally or deployed on a server, enabling lightweight access and quick demonstrations.
The Tkinter modality provides an offline desktop GUI suited for local environments.
Since inference is performed entirely on CPU, deployment is straightforward and does not require GPU drivers or specialized hardware.

To facilitate inspection and reuse, upon publication of the manuscript we will publicly release the complete package of software and tutorials through the GitHub and Kaggle repositories specified in this work. 

\section{Character Segmentation}
\label{sec3}

\begin{figure}[!b]
	\centering
	\includegraphics[width=0.5\linewidth]{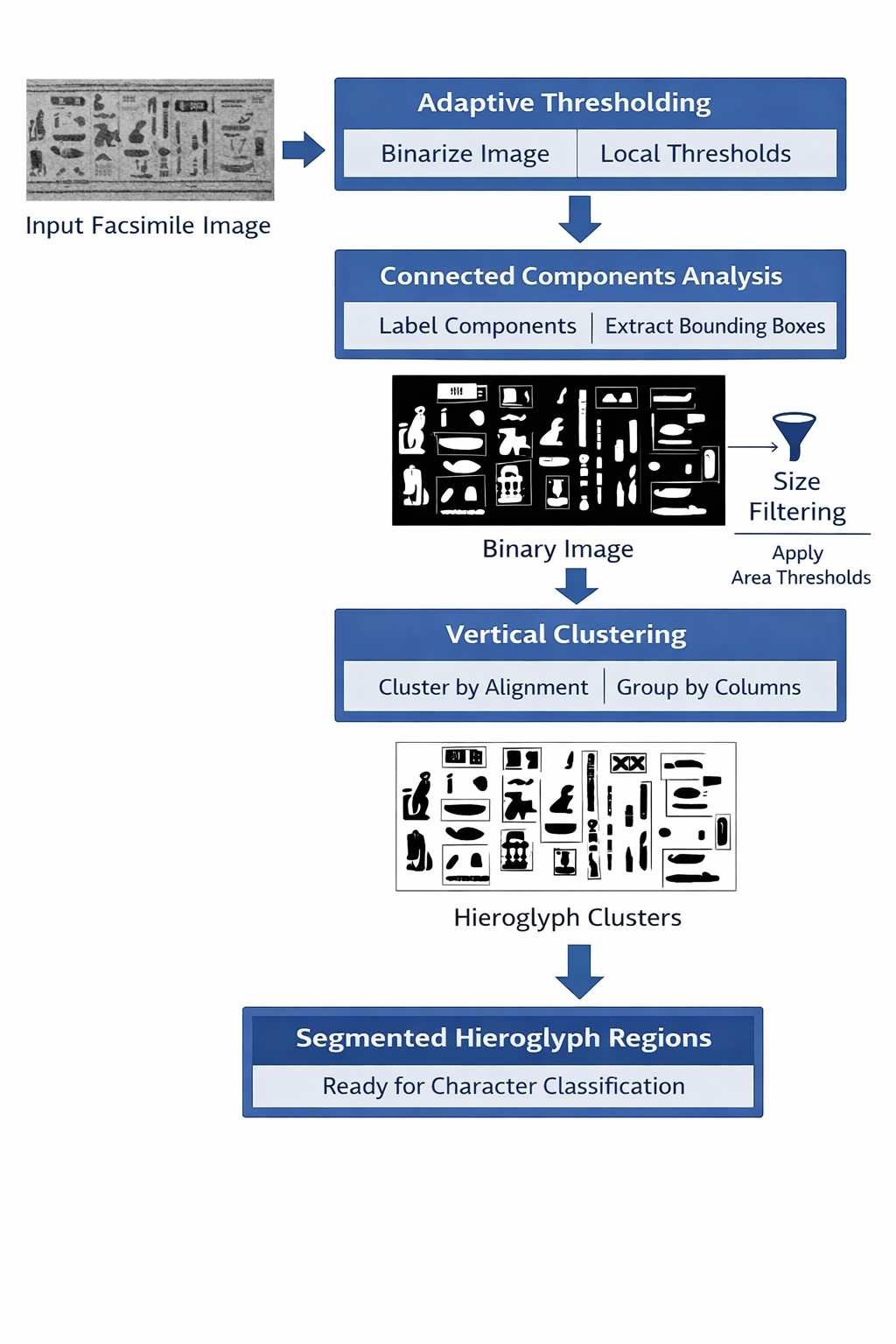} 
	\caption{Proposed transcription workflow using facsimiles as input.}\label{fig:segmentation}
\end{figure}

This preprocessing block involves several stages, as can be observed in Fig.~\ref{fig:segmentation}, designed to identify and segment distinct hieroglyph  regions from the source images of facsimiles in \cite{Buck61}, which are then used for character classification.
Firstly, the input facsimile image is binarized using an adaptive thresholding technique to reduce the impact from variations in illumination or background texture. This method calculates thresholds locally across the image, effectively separating foreground elements (potential hieroglyphs) from the background. Secondly, a Connected Components Analysis (CCA) is applied to the binary image, identifying contiguous regions of foreground pixels. Each resulting connected component represents a potential hieroglyph, and bounding boxes are extracted for each component. To filter out irrelevant regions, components whose area or dimensions fall below a minimum threshold (likely corresponding to noise or minor marks) or exceed a maximum threshold (potentially corresponding to large illustrations, borders, or multiple merged symbols) are discarded. These thresholds are determined empirically, based on the expected scale of hieroglyphs within the facsimiles. Finally, since the Coffin texts are arranged vertically in \cite{Buck61} (see section~\ref{sec2_1}), a clustering step is performed on the centroids or bounding boxes of the remaining candidate hieroglyphs. By identifying groups of components that exhibit strong vertical alignment, we can infer the presence of distinct columns of text. This step aids in organizing the extracted hieroglyphs and potentially assists in determining reading order.

\section{Character Classification}
\label{sec4}

For classifying segmented hieroglyphs into Manuel de Codage (MdC) categories, we propose and compare three different strategies hereinafter: (1) \textbf{Traditional Machine Learning}, using handcrafted features (e.g., HOG) with an SVM classifier; (2)\textbf{End-to-end Deep classification}, fine-tuning a CNN directly with a weighted loss function; and (3) \textbf{Deep metric learning}, where a CNN generates feature embeddings for similarity-based classification against class prototypes.  Our primary method is the proposed Deep Metric Learning approach. For benchmarking purposes, we additionally implement the other methods Detailed descriptions of each method are provided next.

\subsection{Traditional Machine Learning (Classifiers)}

As a comparative baseline to the deep metric learning approaches proposed before, a traditional machine learning pipeline has also been implemented and assessed. This method extracts handcrafted visual features from the hieroglyph images, followed by training a standard supervised classifier.

The feature extraction collects a set of predefined characteristics from each greyscale hieroglyph image $I$. The feature vector $\mathbf{x}$ for each image $I$ is formed by concatenating the following descriptors:

\begin{itemize}
	\item \textbf{Histogram of Oriented Gradients (HOG):} The HOG features~\cite{dalal2005} are obtained to capture the local shape and edge information. We use $12$ orientation bins, $5 \times 5$ pixels per cell, and $1 \times 1$ cells per block, with square-root normalization. The resulting HOG descriptor vector is normalized by its maximum absolute value.
	
	\item \textbf{Projection Profiles:} Normalized sums of pixel intensities are calculated along the horizontal (X-axis) and vertical (Y-axis) directions. These profiles provide a global representation of the symbol's mass distribution.
	
	\item \textbf{Diagonal Projections:} To capture information along diagonal axes, pixel intensities are summed along the main diagonal (top-left to bottom-right) and the secondary diagonal (top-right to bottom-left) across various offsets. These sums are also normalized.
\end{itemize}

The individual feature components, normalized HOG, horizontal projection, vertical projection, main diagonal projection and secondary diagonal projection, are concatenated to form a single feature vector $\mathbf{x} \in \mathbb{R}^m$ for every hieroglyph image $I$, where $m$ is the total dimensionality of the combined features.

\begin{figure}[!b]
	\centering
	\includegraphics[width=0.99\linewidth]{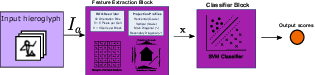} 
	\caption{Example workflow of the proposed classical approximation called as ``Traditional Machine Learning''.}\label{fig_9}
\end{figure}

After extracting the handcrafted feature vectors $\mathbf{x}$, we classify the hieroglyphs using a Support Vector Machine (SVM) classifier~\cite{cortes1995} with a linear kernel. This choice of kernel and hyperparameters was supported by a parametric sweep of the main SVM hyperparameters, selecting the best configuration based on performance on the validation set. In all experiments, the data partitioning follows the predefined training and validation splits reported in Table~\ref{tab:dataset}, rather than a random split, to ensure consistency across methods. Given the pronounced class imbalance, we use class weights computed from the training distribution to promote the most balanced weighting possible, thereby penalizing misclassifications of minority classes more heavily and avoiding a bias towards frequent symbols. The final linear SVM model is trained on the training set and assessed on the corresponding evaluation split(s) as described in Table~\ref{tab:dataset}. We show in Figure \ref{fig_9} the workflow of the proposed metric learning architecture, once trained.

\subsection{End-to-End Deep CNN (Classifier)}

Furthermore, an additional approach has been considered for comparison's sake, where a pre-trained CNN network is fine-tuned as an end-to-end hieroglyph classifier. 
Similarly to the feature extractor previously used in the metric learning approach in Section \ref{sec4_1}, the MobileNetV2 architecture \cite{sandler2018}, pre-trained on ImageNet \cite{deng2009}, has been adapted and directly applied to hieroglyph identification. For that purpose, the original MobileNet's classification head has been modified by adding a flatten layer, a fully-connected layer with $512$ neurons and a rectified linear activation, and a final fully-connected layer, with a number of neurons equal to the number of unique MdC hieroglyph classes $C$ to be identified in our training dataset and a softmax activation function.

This model has been trained by using a standard supervised learning. Since the inherent class imbalance in the dataset remains, where some symbols appear far more frequently than others, a simple categorical cross-entropy loss might lead to a model biased towards predicting common classes. To mitigate this, a weighted categorical cross-entropy loss function has been considered. Every class $c$ is assigned a weight $w_c$, typically obtained as inversely proportional to its frequency in the training set. The loss for a batch of $N$ samples is then computed as:

\begin{equation} \label{eq:weighted_cce}
	\mathcal{L} = -\frac{1}{N} \sum_{i=1}^N \sum_{c=1}^C w_c \cdot y_{ic} \cdot \log(p_{ic})
\end{equation}

where $y_{ic}$ is the one-hot encoded true label (1 if sample $i$ belongs to class $c$, 0 otherwise), and $p_{ic}$ is the model's predicted confidence for sample $i$ belonging to class $c$. This weighting scheme increases the penalty for misclassifying samples from less frequent classes, encouraging the model to learn representations that are effective across the entire range of hieroglyphs.

\begin{figure}[!b]
	\centering
	\includegraphics[width=0.90\linewidth]{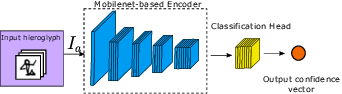} 
	\caption{Example workflow of the proposed End-to-End Deep CNN architecture.}\label{fig_8}
\end{figure}

We show in Figure \ref{fig_8} the workflow of the proposed metric learning architecture, once trained.

\subsection{Deep Metric Learning (Classifiers and Clustering)}
\label{sec4_1}
This approach, based on metric learning, tries to create a feature-based ancient Egyptian hieroglyph classification system. The main motivation of this method is to address the challenges posed by datasets with small support and highly unbalanced classes. The methodology consists of two main steps: feature extraction using a fine-tuned deep neural network and classification based on similarity in the learned feature space.

As for the deep feature extraction, a deep Convolutional Neural Network (CNN) architecture based on MobileNetV2~\cite{ sandler2018} is considered as the backbone for feature extraction. The model is pre-trained on the large-scale ImageNet dataset \cite{deng2009}, thus capturing hierarchical visual representations learned by the network on diverse images. This pre-trained model has been adapted to this specific task by removing its original top classification layers. The output feature maps from the MobileNetV2 base backbone are then flattened into a one-dimensional vector. This high-dimensional vector is subsequently passed through a fully connected layer to reduce its dimensionality to $d=128$. The output of this dense layer is L2-normalized, thus ensuring that all the extracted feature vectors lie on the surface of a $d$-dimensional unit hypersphere. This normalization is beneficial for distance metrics, such as cosine similarity. The whole network, including the MobileNetV2 base, is then fine-tuned during the following metric learning phase, described here below.



Formally, let $I$ be an input hieroglyph image. After standard preprocessing (resizing to a fixed input size and normalization), the image is passed through the modified CNN, denoted by the function $f(\cdot)$. This function maps the image $I$ to a dense feature vector or embedding $\mathbf{z}$:

\begin{equation}
	\mathbf{z} = f(I) \in \mathbb{R}^d
	\label{eq:features}
\end{equation}

where $d$ is the dimensionality of the embedding space and $\|\mathbf{z}\|_2 = 1$. 

With regard to the metric learning step, the core idea is to learn an embedding space where images of the same hieroglyph (same MdC code) are mapped to nearby points, whereas images from different hieroglyphs are mapped to distant points. This is achieved by training or fine-tuning the feature extractor $f(\cdot)$ with objectives characteristic of metric learning, such as minimizing the intra-class distance and maximizing the inter-class distance within the embedding space $\mathbb{R}^d$.

To optimize the embedding space, a Siamese network architecture is trained with a contrastive loss objective based on cosine similarity. The Siamese network consists of two identical branches, each composed of the base feature extraction network $f(\cdot)$ defined above, sharing the same weights. During training, the network is presented with pairs of hieroglyph images $(I_a, I_b)$. These pairs are constructed dynamically: positive pairs consist of two different images belonging to the same MdC class, whereas negative pairs consist of two images from different MdC classes. To enhance robustness and generalization, significant data augmentation is applied exclusively during the generation of training pairs, and it is not used for validation or testing. Augmentation is applied to a given sample with a probability of 50\%. When triggered, it includes standard geometric transformations, namely small rotations limited to $\pm 15^{\circ}$ and small width/height shifts, as well as optional mirroring \emph{only} along the horizontal axis. This restriction is adopted because vertical flips and unrestricted mirroring may generate sign configurations that do not correspond to valid hieroglyphs, potentially introducing unrealistic training examples. In addition, we introduce synthetic artefacts such as random bands along image borders and circular occlusions. These occlusions are intended to emulate imperfections commonly found in practice, either due to variability in expert drawings or to manuscript degradation over time; in binarized facsimiles, such effects often manifest as darkened regions, missing strokes, or partially damaged symbols.
The objective of the training is to tune the weights of $f(\cdot)$ such that the cosine similarity between the embeddings $(\mathbf{z}_a, \mathbf{z}_b) = (f(I_a), f(I_b))$ is high (close to $1$) for positive pairs, and low (pushed below a margin $m$) for negative pairs. This is achieved by minimizing a cosine contrastive loss function $L$. For a certain pair with a predicted cosine similarity $s = \mathbf{z}_a \cdot \mathbf{z}_b$ and a ground-truth label $y$ ($y=0$ for similar, $y=1$ for dissimilar), the loss is defined as:

\begin{equation} \label{eq:contrastive_loss}
	L(s, y) = (1-y)(1-s)^2 + y \cdot \max(s - m, 0)^2
\end{equation}

where $m$ is a predefined margin hyperparameter (e.g., $m=0.5$). The network is trained using the Adam optimizer \cite{kingma2014} with standard training procedures, including mini-batch gradient descent, early stopping based on validation loss, and learning rate reduction. This process fine-tunes the feature extractor $f(\cdot)$ to produce embeddings that effectively cluster images from the same hieroglyph class, while separating images from different classes in the cosine similarity space.

Finally, after the feature extractor $f(\cdot)$ is trained, a template-based representation is established for classification. For each known MdC class $k$ present in our training dataset $S$, a representative template vector is computed, as well as a centroid $\mathbf{c}_k$, by averaging the L2-normalized embeddings of all the training examples belonging to that class, according to:

\begin{equation}
	\mathbf{c}_k = \frac{1}{|S_k|} \sum_{I_i \in S_k} f(I_i)
\end{equation}

where $S_k = \{I_i \in S \mid \text{label}(I_i) = k\}$ is the set of training images for class $k$. 
In order to recognize an unknown hieroglyph image $I_{new}$, its feature embedding $\mathbf{z}_{new} = f(I_{new})$ is firstly obtained by using the fine-tuned feature extractor. Then, this new embedding is compared to all the pre-computed class centroids $\{\mathbf{c}_k\}$ using the cosine similarity defined in \eqref{eq:cos_sim}.

\begin{equation}
	\label{eq:cos_sim}
	\text{similarity}(\mathbf{z}_{new}, \mathbf{c}_k) = \frac{\mathbf{z}_{new} \cdot \mathbf{c}_k}{\|\mathbf{z}_{new}\| \|\mathbf{c}_k\|} = \frac{\mathbf{z}_{new} \cdot \mathbf{c}_k}{\|\mathbf{c}_k\|} \quad (\text{since } \|\mathbf{z}_{new}\| = 1)
\end{equation}
The predicted MdC code $\hat{k}$ for the input image $I_{new}$ is assigned to the class whose centroid yields the highest cosine similarity \eqref{eq:k_est}. 
This approach performs the nearest centroid classification in the learned embedding space, effectively using the discriminative power of the fine-tuned deep features.

\begin{equation}
	\label{eq:k_est}
	\hat{k} = \underset{k}{\operatorname{argmax}} \left( \text{similarity}(\mathbf{z}_{new}, \mathbf{c}_k) \right)
\end{equation}

\begin{figure}[!t]
	\centering
	\includegraphics[width=0.70\linewidth]{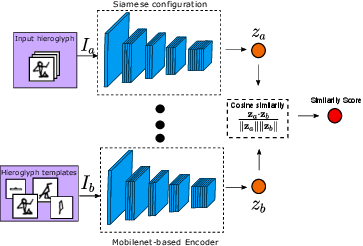} 
	\caption{Example workflow of the proposed Deep Metric Learning architecture.}\label{fig_7}
\end{figure}

We show in Figure \ref{fig_7} the workflow of the proposed metric learning architecture, once trained.

\section{Experimental Results}
\label{sec5}

This section compares the performance of the three classification methods detailed in Section~\ref{sec4}. For clarity, the short names that will be used in the following sections and tables are introduced here: (1) our proposed Deep Metric Learning model (\method{Deep-MML}); (2) the Traditional Machine Learning baseline (\method{Trad-ML}); and (3) the End-to-End Deep CNN Classifier (\method{CNN-End2End}). 


As discussed in Section~\ref{sec2_1}, the available corpus exhibits significant class imbalances, which poses a considerable challenge to the learning methods. To assess this, we conducted two experiments focusing on subsets of hieroglyphs (classes): one with $140$ classes and another with $222$ classes. These subsets include hieroglyphs with a minimum support of $25$ and $2$ samples per class, respectively. Table~\ref{tab1} presents the balanced accuracy achieved by each method on two distinct evaluation sets:
\begin{itemize}
	\item The ``Test 15\%'' set comprises a random aproximately 15\% split of the labeled and pre-segmented hieroglyphs, with the remaining 61\% used for training and another 15\% for validation.
	\item The ``Test 10 pages'' set consists of $10$ complete pages from the corpus, which were held out and reserved exclusively for testing, as described in Section~\ref{sec2_1}. The creation of this evaluation set guarantees that no information from these ten pages is used during the training or validation of the system. For this set, hieroglyphs first required segmentation using the method described in Section~\ref{sec2}.

\end{itemize}

In the $140$-class experiment, the \method{Deep-MML} and \method{CNN-End2End} methods achieved similar balanced accuracy on the ``Test 15\%'' set, with \method{Deep-MML} performing slightly better at $97.70\%$ compared to $96.20\%$ for \method{CNN-End2End}. The \method{Trad-ML} obtains a competitive accuracy of $93.50\%$. However, on the ``Test 10 pages'' set, which requires segmentation and tests generalization more rigorously, a drop in accuracy was observed for all methods, and the performance differences became more pronounced. Here, \method{Deep-MML} maintained the highest accuracy at $93.5\%$, followed by \method{Trad-ML} at $88.43\%$. The accuracy of \method{CNN-End2End} decreased to $84.20\%$. These results suggest that the ``Test 10 pages'' set effectively highlights the generalization capabilities of the methods, with \method{Deep-MML} demonstrating superior performance in this scenario. 

The $222$-class experiment further highlights the sensitivity of the \method{Trad-ML} and \method{CNN-End2End} methods to highly imbalanced classes. The balanced accuracy of \method{Trad-ML} dropped to $89.64\%$ on the ``Test 15\%'' set and further to $78.81\%$ on the ``Test 10 pages'' set. The performance of \method{CNN-End2End} declined dramatically, reaching $74.55\%$ on the ``Test 15\%'' set and falling below $60\%$ ($58.75\%$) on the ``Test 10 pages'' set. In contrast, \method{Deep-MML} consistently outperformed the other methods, maintaining reasonably high accuracy across both test sets in this challenging experiment. It achieved $95.90\%$ on the ``Test 15\%'' set and $88.9\%$ on the ``Test 10 pages'' set. Given its consistently higher balanced accuracy, especially in the held-out pages setting, Deep-MML is the classifier adopted in the final OCR-PT-CT system.
\begin{table}[htbp]
	\centering
	\resizebox{\textwidth}{!}{%
		\begin{tabular}{lcccc}
			\toprule
			\textbf{Approach} &   \textbf{Test 15\%} (140-class) & \textbf{Test 15\%} (222-class) & \textbf{Test 10 pages} (140-class) & \textbf{Test 10 pages} (222-class) \\
			\midrule
			\method{Deep-MML} & \textbf{97.70}\% & \textbf{95.90\%} &\textbf{93.50\%}& \textbf{88.90\%} \\
			\method{Trad-ML} &   92.46\% &89.64\%& 88.43\%& 78.81\%\\
			\method{CNN-End2End} & 93.87\% &70.70\%& 83.93\%& 54.79\%\\
			\bottomrule
		\end{tabular}%
	}
	\caption{Overall balanced accuracy performance of the proposed classification methods. Results are shown for experiments with $140$ and $222$ classes on the ``Test 15\%'' and ``Test 10 pages'' evaluation sets.}
	\label{tab1}
\end{table}
For a more detailed analysis, Table~\ref{tab3} presents the F1-score per major MdC category (denoted by letters A-Z) for the \method{Deep-MML} and \method{CNN-End2End} classifiers. Each category includes a group of individual hieroglyph classes from the dataset. Generally, categories with a high number of training samples per hieroglyph were classified well by both methods, particularly in the $140$-class experiment. However, when expanding to $222$ classes, the performance of \method{CNN-End2End} deteriorated, mainly because many categories included hieroglyphs with very few samples. An extreme example is category C, which contains no hieroglyphs with more than $25$ samples (hence its absence in the $140$-class results for this category). As shown, \method{CNN-End2End} produced one of its lowest F1-scores for category C in the $222$-class experiment, likely due to severe class imbalance. Conversely, \method{Deep-MML} demonstrated more consistent performance across different categories and for both the $140$-class and $222$-class experiments.

\begin{longtable}[c]{|m{2cm}|m{2.7cm}|m{2.7cm}|m{2cm}|m{2cm}|}
	\caption{F1-score per hieroglyph group obtained by the fine-tuned CNN approach and the deep metric learning for $140$ and $222$ classes.} \\
	\hline
	Hieroglyph group & CNN-End2End ($140$ classes) & CNN-End2End ($222$ classes) & Deep-MML ($140$ classes) & Deep-MML ($222$ classes) \\ \hline
	\endfirsthead
	
	A & 0.89 & 0.54 & 0.98 & 0.95 \\ \hline
	B & 0.96 & 0.27 & 0.72   & 0.90 \\ \hline
	C & - & 0.31 & -   &   1.0 \\ \hline
	D & 0.96 & 0.59 & 0.98   &   0.94 \\ \hline
	E & 0.94 & 0.53 & 1.0   &   0.76 \\ \hline
	F & 0.91 & 0.59 & 0.98    & 0.99 \\ \hline
	G & 0.91 & 0.69 & 1.0   & 0.99 \\ \hline
	H & 0.90 & 0.57 & 1.0   &   1.0 \\ \hline
	I & 0.97 & 0.87 & 1.0   &   1.0 \\ \hline
	K & 0.96 & 0.27 & 1.0   &   0.4 \\ \hline
	M & 0.91 & 0.66 & 0.98    & 0.99 \\ \hline
	N & 0.95 & 0.66 & 0.99   & 0.98 \\ \hline
	O & 0.91 & 0.64 & 0.94   & 1.0 \\ \hline
	P & 0.97 & 0.67 & 1.0   &   1.0 \\ \hline
	Q & 0.97 & 0.42 & 1.0   & 0.99 \\ \hline
	R & 0.93 & 0.84 & 0.99    & 1.0 \\ \hline
	S & 0.95 & 0.58 & 1.0    & 0.97 \\ \hline
	T & 0.80 & 0.51 & 0.89    & 0.79 \\ \hline
	U & 0.98 & 0.69 & 1.0    &  1.0 \\ \hline
	V & 0.93 & 0.53 & 0.95   & 0.99 \\ \hline
	W & 0.96 & 0.60 & 1.0    &  1.0 \\ \hline
	X & 0.86 & 0.74 & 0.92 & 1.0 \\ \hline
	Y & 0.97 & 0.71 & 1.0 & 1.0 \\ \hline
	Z & 0.94 & 0.66 & 0.95   &   0.83 \\ \hline
\label{tab2}
\end{longtable}

Finally, Table \ref{tab3} includes the performance (Precision, Recall and F1-score) for every single hieroglyph class considered in the $140$-class experiment. As expected, hieroglyphs supported by a lower number of samples present a degraded classification performance, such as for the class O29 or T3, whereas those with a high number of samples available are correctly classified by both \method{Deep-MML} and \method{CNN-End2End} classifiers.

	\footnotesize
	\begin{longtable}{|c|c|c|c|c|c|c|c|}
		\caption{Disaggregated performance per hieroglyph obtained by the fine-tuned CNN approach and the deep learning metric one for a set of $140$ classes.} \\
		\hline
		Hieroglyph & \multicolumn{3}{c|}{CNN-End2End} & \multicolumn{3}{c|}{Deep-MML} & Support \\
		& Precision & Recall & F1-score  & Precision& Recall&F1-score&  \\
		\hline
		\endfirsthead
		
		A1 & 1.00 & 0.98 & 0.99 &1.00&1.00&1.00& 93 \\ \hline
		A2 & 1.00 & 0.69 & 0.81 &1.00&1.00&1.00& 16 \\ \hline
		A24 & 1.00 & 0.87 & 0.93 &1.00&1.00&1.00& 15 \\ \hline
		A40 & 0.94 & 0.97 & 0.96 &0.68&0.82&0.74& 124 \\ \hline
		A49 & 0.93 & 1.00 & 0.97 &0.8&1.0&0.89& 14 \\ \hline
		A50 & 1.00 & 0.96 & 0.98 &1.00&1.00&1.00& 268\\ \hline
		Aa1 & 1.00 & 1.00 & 1.00 &1.00&1.00&1.00& 243 \\ \hline
		Aa15 & 0.65 & 1.00 & 0.79    &1.00&1.00&1.00& 13 \\ \hline
		Aa2 & 0.81 & 0.89 & 0.85    &1.00&1.00&1.00& 19 \\ \hline
		Aa21 & 1.00 & 1.00 & 1.00    &1.00&1.00&1.00& 13 \\ \hline
		Aa27 & 0.62 & 1.00 & 0.76    &1.00&1.00&1.00&   8 \\ \hline
		Aa8 & 0.50 & 1.00 & 0.67    &1.00&1.00&1.00&   8 \\ \hline
		B1 & 0.97 & 0.95 & 0.96   &0.86&0.73&0.79& 175 \\ \hline
		D1 & 0.98 & 1.00 & 0.99   &1.00&1.00&1.00&   40 \\ \hline
		D2 & 1.00 & 0.96 & 0.98 &1.00&1.00&1.00& 295 \\ \hline
		D21 & 0.97 & 0.99 & 0.98 &1.00&1.00&1.00& 718 \\ \hline
		D24 & 1.00 & 1.00 & 1.00   &1.00&1.00&1.00&    8 \\ \hline
		D35 & 1.00 & 0.89 & 0.94   &1.0&1.00&1.00&    9 \\ \hline
		D36 & 1.00 & 0.77 & 0.87   &1.00&0.98&0.99& 155 \\ \hline
		D37 & 1.00 & 1.00 & 1.00   &0.93&0.87&0.90&   85 \\ \hline
		D4 & 0.99 & 1.00 & 0.99 &1.00&1.00&1.00& 216 \\ \hline
		D40 & 0.85 & 0.98 & 0.91  &0.76&0.90&0.82&    46 \\ \hline
		D46 & 1.00 & 0.91 & 0.95 &0.99&1.00&0.99& 234 \\ \hline
		D50 & 0.80 & 0.92 & 0.86    &0.89&1.00&0.94& 13 \\ \hline
		D52 & 0.98 & 1.00 & 0.99   &1.00&1.00&1.00&   50 \\ \hline
		D53 & 1.00 & 0.94 & 0.97    &1.00&1.00&1.00& 18 \\ \hline
		D55 & 1.00 & 0.98 & 0.99 &1.00&1.00&1.00& 207 \\ \hline
		D56 & 1.00 & 1.00 & 1.00    &1.00&1.00&1.00&  31 \\ \hline
		D58 & 0.99 & 0.98 & 0.98&1.00&1.00&1.00& 231 \\ \hline
		E23 & 0.95 & 1.00 & 0.98   &1.00&1.00&1.00&   20 \\ \hline
		E34 & 0.82 & 1.00 & 0.90    &1.00&1.00&1.00& 14 \\ \hline
		F13 & 1.00 & 1.00 & 1.00    &1.00&1.00&1.00& 15 \\ \hline
		F18 & 0.18 & 1.00 & 0.30    &0.88&1.00&0.933& 10 \\ \hline
		F21 & 0.95 & 1.00 & 0.98   &1.00&0.91&0.95&   20 \\ \hline
		F27 & 1.00 & 1.00 & 1.00    &1.00&1.00&1.00& 15 \\ \hline
		F32 & 1.00 & 1.00 & 1.00    &0.90&0.90&0.90& 18 \\ \hline
		F35 & 1.00 & 1.00 & 1.00    &1.00&1.00&1.00&  20 \\ \hline
		F4 & 1.00 & 1.00 & 1.00    &1.00&1.00&1.00&   9 \\ \hline
		F40 & 1.00 & 1.00 & 1.00    &1.00&1.00&1.00& 16 \\ \hline
		F51 & 0.91 & 1.00 & 0.95   &1.00&1.00&1.00&   21 \\ \hline
		G1 & 1.00 & 0.70 & 0.83   &1.00&1.00&1.00& 115 \\ \hline
		G17 & 0.99 & 0.98 & 0.99 &1.00&0.98&0.99& 200 \\ \hline
		G17+M17 & 1.00 & 1.00 & 1.00   &1.00&1.00&1.00&    8 \\ \hline
		G190 & 0.32 & 1.00 & 0.49   &1.00&1.00&1.00&   25 \\ \hline
		G36 & 0.97 & 0.95 & 0.96  &1.00&1.00&1.00&    38 \\ \hline
		G39 & 0.99 & 0.98 & 0.98  &1.00&1.00&1.00&    82 \\ \hline
		G4 & 1.00 & 1.00 & 1.00    &1.00&1.00&1.00& 10 \\ \hline
		G43 & 1.00 & 1.00 & 1.00  &1.00&1.00&1.00&    34 \\ \hline
		G5 & 0.89 & 1.00 & 0.94  &0.91&1.00&0.95&    34 \\ \hline
		H38 & 1.00 & 0.73 & 0.84   &1.00&1.00&1.00&   33 \\ \hline
		H39 & 0.95 & 0.91 & 0.93   &1.00&1.00&1.00&   23 \\ \hline
		H40 & 1.00 & 0.67 & 0.80    &1.00&1.00&1.00&   9 \\ \hline
		H6 & 0.95 & 0.95 & 0.95    &1.00&1.00&1.00&  20 \\ \hline
		H6A & 1.00 & 0.85 & 0.92    &1.00&1.00&1.00& 13 \\ \hline
		H8 & 0.91 & 0.98 & 0.94    &0.92&1.00&0.96&  41 \\ \hline
		I10 & 1.00 & 0.98 & 0.99   &1.00&1.00&1.00&   44 \\ \hline
		I9 & 1.00 & 0.92 & 0.96   &1.00&1.00&1.00&   94 \\ \hline
		K1 & 1.00 & 0.92 & 0.96   &1.00&1.00&1.00&   26 \\ \hline
		M16 & 1.00 & 0.91 & 0.95    &1.00&1.00&1.00& 11 \\ \hline
		M17 & 0.98 & 0.96 & 0.97 &1.00&1.00&1.00& 363 \\ \hline
		M18 & 1.00 & 0.90 & 0.95  &1.00&0.92&0.96&    21 \\ \hline
		M22 & 1.00 & 0.94 & 0.97    &1.00&1.00&1.00& 17 \\ \hline
		M23 & 1.00 & 0.96 & 0.98   &1.00&1.00&1.00&   24 \\ \hline
		M29 & 1.00 & 0.77 & 0.87    &1.00&1.00&1.00& 13 \\ \hline
		M37 & 1.00 & 0.66 & 0.79    &1.00&1.00&1.00&  32 \\ \hline
		M42 & 0.93 & 0.93 & 0.93   &0.94&0.94&0.94&   29 \\ \hline
		M44 & 0.75 & 0.75 & 0.75   &1.00&1.00&1.00&    8 \\ \hline
		N1 & 1.00 & 1.00 & 1.00   &1.00&1.00&1.00& 147 \\ \hline
		N18 & 0.73 & 1.00 & 0.84    &1.00&1.00&1.00& 16 \\ \hline
		N23 & 1.00 & 0.97 & 0.98   &1.00&1.00&1.00&   99 \\ \hline
		N25 & 0.96 & 1.00 & 0.98   &0.97&1.00&0.98&   53 \\ \hline
		N26 & 0.93 & 0.93 & 0.93   &1.00&0.94&0.97&   30 \\ \hline
		N27 & 1.00 & 1.00 & 1.00  &1.00&1.00&1.00&    24 \\ \hline
		N29 & 0.83 & 1.00 & 0.91   &1.00&1.00&1.00& 101 \\ \hline
		N31 & 0.97 & 1.00 & 0.99   &1.00&1.00&1.00&   35 \\ \hline
		N35 & 1.0 & 0.99 & 0.99 &1.00&1.00&1.00& 641 \\ \hline
		N37 & 1.00 & 0.89 & 0.95 &1.00&0.99&0.99& 204 \\ \hline
		N38 & 0.80 & 1.00 & 0.89    &1.00&1.00&1.00&12 \\ \hline
		N40 & 1.00 & 0.90 & 0.95    &1.00&1.00&1.00& 10 \\ \hline
		N42 & 0.90 & 0.98 & 0.94   &1.00&0.97&0.99&   62 \\ \hline
		N5 & 1.00 & 1.00 & 1.00   &1.00&1.00&1.00&   53 \\ \hline
		O1 & 1.00 & 0.99 & 0.99  &1.00&1.00&1.00& 124 \\ \hline
		O28 & 0.85 & 0.94 & 0.89    &1.00&1.00&1.00& 18 \\ \hline
		O29 & 0.46 & 0.79 & 0.58    &0.89&1.00&0.94& 14 \\ \hline
		O31 & 0.91 & 1.00 & 0.95   &1.00&1.00&1.00&   30 \\ \hline
		O34 & 1.00 & 0.99 & 0.99   &1.00&0.99&0.99& 194 \\ \hline
		O4 & 1.00 & 1.00 & 1.00    &1.00&1.00&1.00&   7 \\ \hline
		O42 & 0.92 & 1.00 & 0.96    &1.00&1.00&1.00& 12 \\ \hline
		O47 & 0.69 & 0.76 & 0.73    &0.71&0.45&0.56&  21 \\ \hline
		O49 & 1.00 & 1.00 & 1.00    &1.00&1.00&1.00&  21 \\ \hline
		O6 & 1.00 & 0.95 & 0.97   &1.00&1.00&1.00&   20 \\ \hline
		P1 & 0.95 & 1.00 & 0.98   &1.00&1.00&1.00&   21 \\ \hline
		P6 & 0.94 & 1.00 & 0.97    &1.00&1.00&1.00& 15 \\ \hline
		P8 & 0.91 & 1.00 & 0.96    &1.00&1.00&1.00& 11 \\ \hline
		Q1 & 0.97 & 0.94 & 0.95   &1.00&1.00&1.00& 179 \\ \hline
		Q3 & 1.0 & 0.98 & 0.99   &1.00&0.99&0.99& 356 \\ \hline
		R14 & 0.92 & 1.00 & 0.96    &1.00&1.00&1.00& 12 \\ \hline
		R17A & 1.00 & 0.82 & 0.90    &1.00&1.00&1.00& 11 \\ \hline
		R4 & 0.91 & 1.00 & 0.95   &1.00&0.96&0.98&   50 \\ \hline
		R8 & 1.00 & 0.81 & 0.90  &1.00&1.00&1.00&   79 \\ \hline
		R8A & 1.00 & 0.90 & 0.95    &1.00&1.00&1.00& 10 \\ \hline
		S2 & 1.00 & 1.00 & 1.00    &1.00&1.00&1.00& 11 \\ \hline
		S24 & 0.93 & 1.00 & 0.96   &1.00&1.00&1.00&  41 \\ \hline
		S29 & 0.99 & 0.99 & 0.99   &1.00&1.00&1.00& 160 \\ \hline
		S33 & 0.78 & 1.00 & 0.87   &1.00&1.00&1.00&  7 \\ \hline
		S34 & 1.00 & 1.00 & 1.00   &1.00&1.00&1.00&   31 \\ \hline
		S42 & 1.00 & 0.98 & 0.99   &1.00&1.00&1.00&   55 \\ \hline
		S43 & 0.75 & 0.94 & 0.83    &1.00&1.00&1.00& 16 \\ \hline
		T21 & 1.00 & 1.00 & 1.00    &1.00&1.00&1.00& 10 \\ \hline
		T22 & 0.89 & 1.00 & 0.94    &1.00&1.00&1.00&   8 \\ \hline
		T28 & 0.53 & 1.00 & 0.69    &1.00&1.00&1.00& 17 \\ \hline
		T3 & 0.43 & 0.67 & 0.52   &0.40&0.33&0.36&    9 \\ \hline
		T30 & 0.73 & 1.00 & 0.84   &0.63&1.00&0.78&   32 \\ \hline
		T34 & 0.87 & 0.78 & 0.82   &1.00&1.00&1.00&    9 \\ \hline
		U15 & 0.96 & 0.96 & 0.96    &1.00&1.00&1.00&  28 \\ \hline
		U2 & 1.00 & 1.00 & 1.00   &1.00&1.00&1.00&  21 \\ \hline
		U23 & 1.00 & 1.00 & 1.00   &1.00&1.00&1.00&    9 \\ \hline
		U29 & 0.95 & 1.00 & 0.97   &1.00&1.00&1.00&  55 \\ \hline
		U33 & 1.00 & 0.97 & 0.9841  &1.00&1.00&1.00&    32 \\ \hline
		V13 & 0.99 & 0.99 & 0.99   &1.00&1.00&1.00& 153 \\ \hline
		V24 & 0.61 & 0.69 & 0.65    &0.64&0.70&0.67& 16 \\ \hline
		V28 & 0.94 & 1.00 & 0.97   &1.00&1.00&1.00&   34 \\ \hline
		V29 & 1.00 & 1.00 & 1.00   &1.00&1.00&1.00&   7 \\ \hline
		V30 & 0.99 & 0.97 & 0.98   &1.00&1.00&1.00& 117 \\ \hline
		V31 & 1.00 & 0.97 & 0.99 &1.00&0.95&0.97& 355 \\ \hline
		V6 & 1.00 & 0.87 & 0.93   &1.00&1.00&1.00&   8 \\ \hline
		W11 & 0.90 & 1.00 & 0.95    &1.00&1.00&1.00&  35 \\ \hline
		W17 & 1.00 & 1.00 & 1.00    &1.00&1.00&1.00&   8 \\ \hline
		W19 & 1.00 & 1.00 & 1.00    &1.00&1.00&1.00& 18 \\ \hline
		W23 & 0.79 & 1.00 & 0.88     &1.00&1.00&1.00& 30 \\ \hline
		W24 & 0.99 & 0.92 & 0.96   &0.98&1.00&0.99& 107 \\ \hline
		W25 & 1.00 & 1.00 & 1.00   &1.00&1.00&1.00&  21 \\ \hline
		X1 & 1.0 & 0.97 & 0.99 &1.00&1.00&1.00&541 \\ \hline
		X2 & 1.00 & 0.95 & 0.97   &1.00&1.00&1.00&   20 \\ \hline
		X4 & 0.54 & 0.50 & 0.52    &0.40&0.67&0.50& 12 \\ \hline
		X8 & 0.96 & 1.00 & 0.98   &1.00&1.00&1.00&   22 \\ \hline
		Y1 & 1.00 & 0.95 & 0.97 &0.99&0.99&0.99& 221 \\ \hline
		Y5 & 0.94 & 1.00 & 0.97   &0.96&1.00&0.98&   44 \\ \hline
		Z1 & 0.91 & 1.00 & 0.95   &1.00&0.96&0.98&   40 \\ \hline
		Z11 & 0.91 & 0.93 & 0.92   &0.94&0.94&0.94&  31 \\ \hline
	\label{tab3}
\end{longtable}
\normalsize

\subsection{5.2. Classifier analysis and operating curves}

In addition to the point estimates reported in Table~\ref{tab2}, we analyze the behaviour of the proposed classifiers as a function of the decision threshold in order to characterize the trade-off between precision and recall and to support the selection of practical operating points. Specifically, we report Precision--Recall (PR) curves and accuracy as a function of the threshold, computed from the models' confidence scores (cosine similarity to the nearest class centroid for Deep-MML, softmax confidence for CNN-End2End). In the multiclass setting, PR curves are obtained in a one-vs-rest manner.

\begin{figure}[!b]
	\centering

	\subfloat{%
		\includegraphics[width=0.49\linewidth]{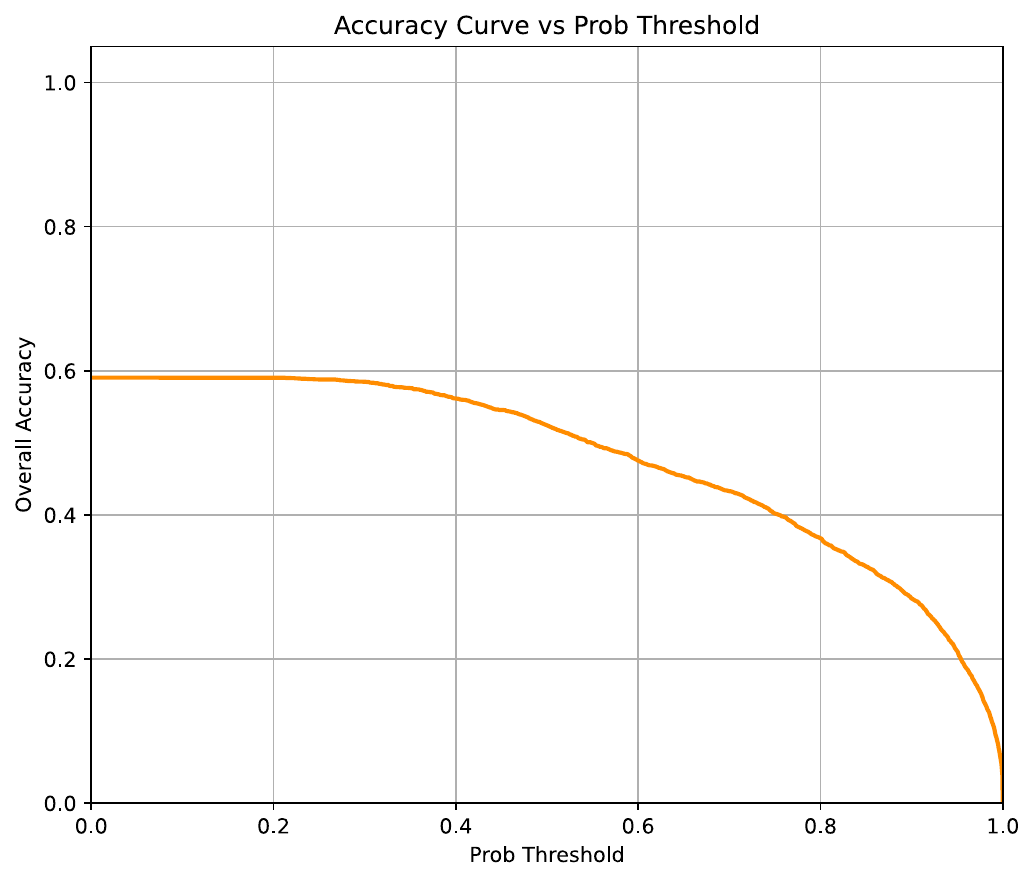}
		
	}
	\subfloat{%
		\includegraphics[width=0.49\linewidth]{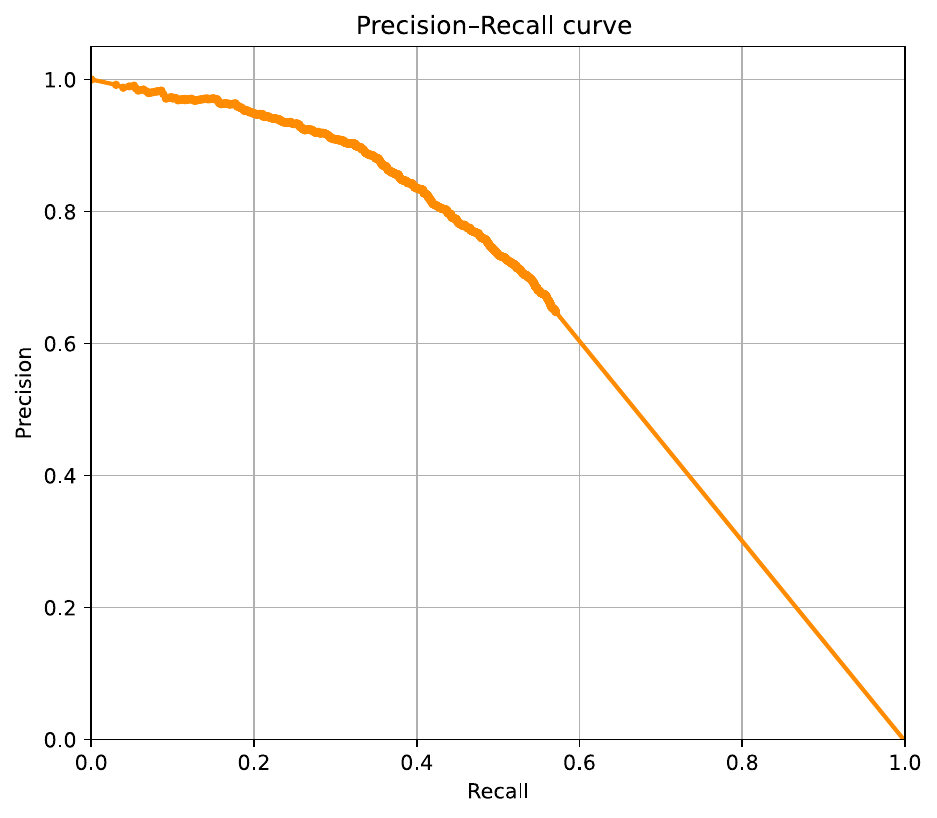}
		
	}

	\vspace{2mm}
	
	\subfloat{%
		\includegraphics[width=0.49\linewidth]{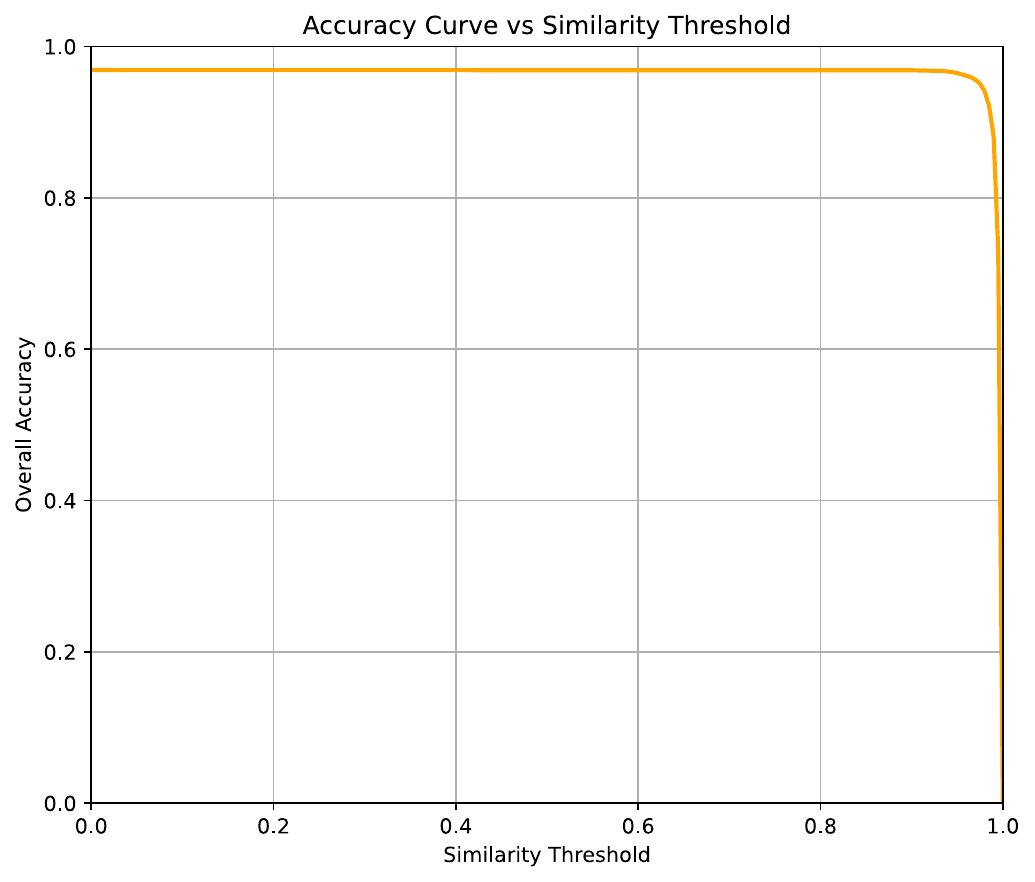}
		
	}
	\subfloat{%
		\includegraphics[width=0.49\linewidth]{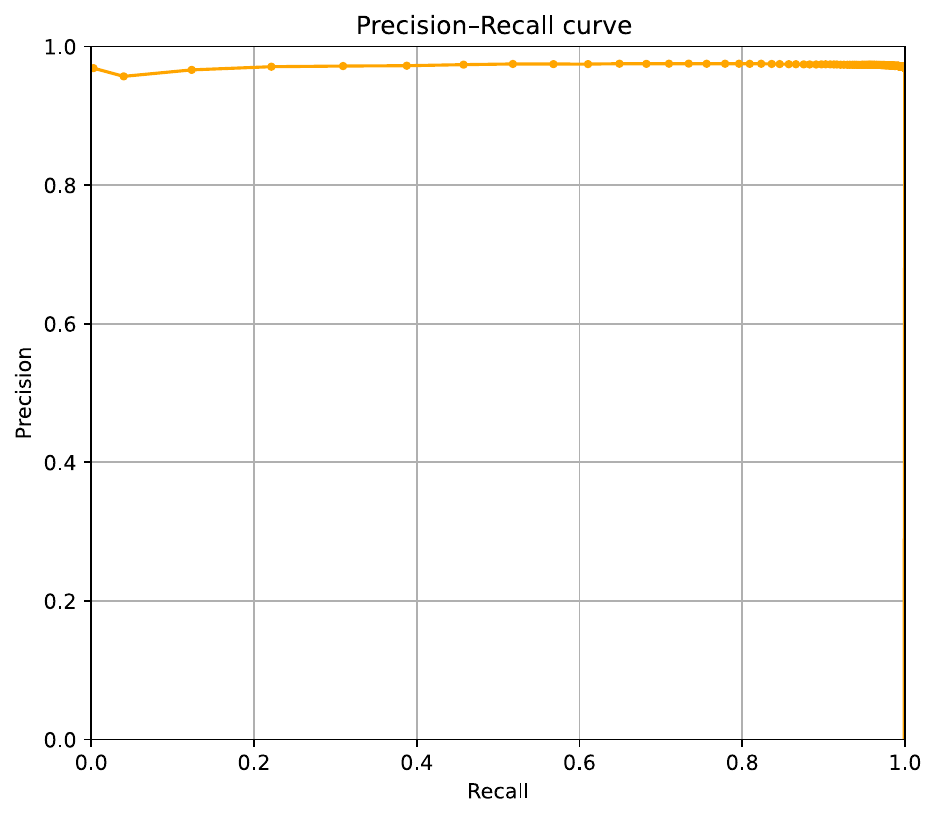}
		
	}
	
	\caption{Accuracy-Threshold and Precision-Recall curves of CNN-End2End being the first row of images and Deep-MML the latter. }
	\label{fig_6}
\end{figure}

All these curves were obtained using 222 classes in the set of 10 isolated pages, representing the most complex case among those considered. The resulting curves are shown in Figure \ref{fig_6}.

In these graphical results, a substantial difference between the two methods is evident, highlighting two key factors. Deep-MML exhibits a higher baseline accuracy than CNN-End2End, which remains very stable independently of the threshold, reaching values close to a 90\% threshold. Furthermore, in the Precision-Recall curves, we observe that the operating point of Deep-MML is highly stable compared to CNN-End2End, making it largely invariant to the choice of threshold. For this analysis, the threshold for CNN-End2End corresponds to the output confidence of the softmax activation, whereas for Deep-MML, it is based on cosine similarity.

\subsection{Study case: Transcription workflow}

\begin{figure}[b]
	\centering
	\includegraphics[width=0.70\linewidth]{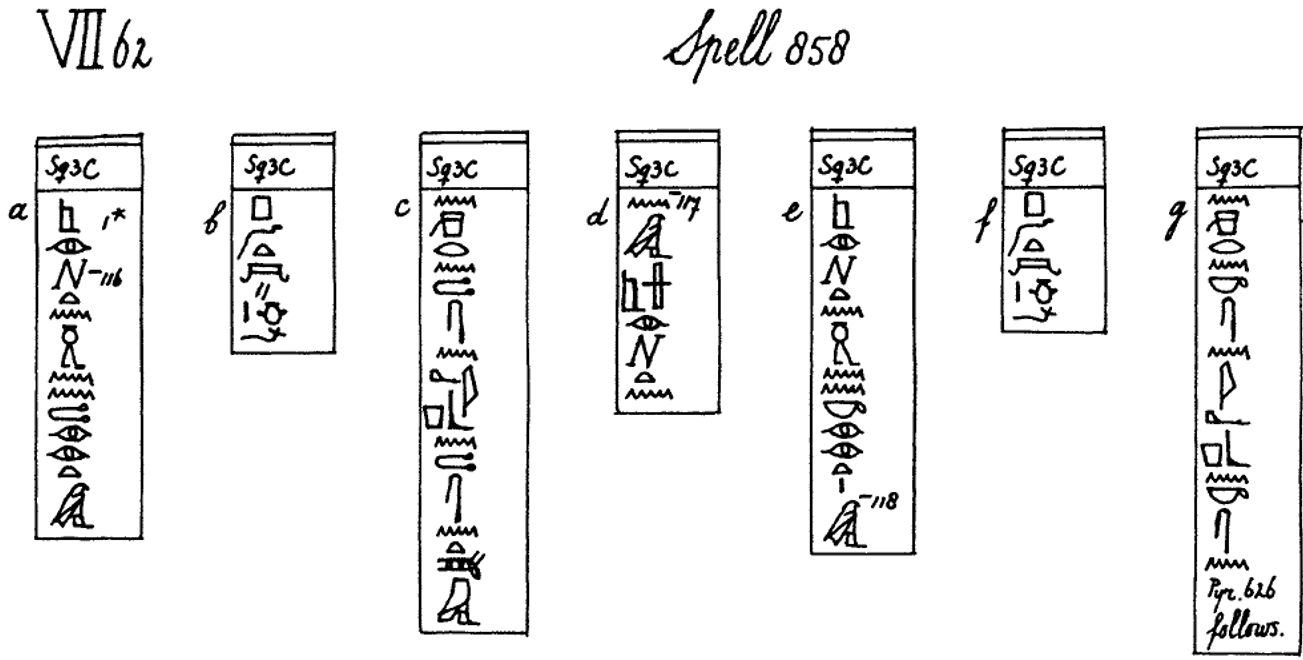} 
	\caption{Ground-truth page corresponding to Spell 858 in \cite{Buck61}, considered as an example.}\label{fig_4}
\end{figure}

The semi-automatic transcription process starts with the loading of the digitized facsimiles from the funerary texts in \cite{Buck61}. For this study case, we use the ground-truth page shown in Fig. \ref{fig_4} as an example. Using a dedicated web application, users can define regions of interest containing columns of hieroglyphs; these regions are first sent to a segmentation module. The segmentation module applies an adaptive process that identifies and isolates individual hieroglyphs. After the cropped images of each hieroglyph are obtained, they are submitted to the classification module, which employs the deep metric learning approach described before. The time required from selection to the visualization of results depends on the number of hieroglyphs analyzed, with an approximate analysis time per hieroglyph of 22 ms on a CPU, using a computer equipped with an AMD Ryzen 5 5600G processor and 16 GB of RAM, which was not fully used. This prediction time per hieroglyph is highlighted because all other processing times are negligible in comparison, being less than 1 ms.

\begin{figure}
	\centering
	\includegraphics[width=0.5\linewidth]{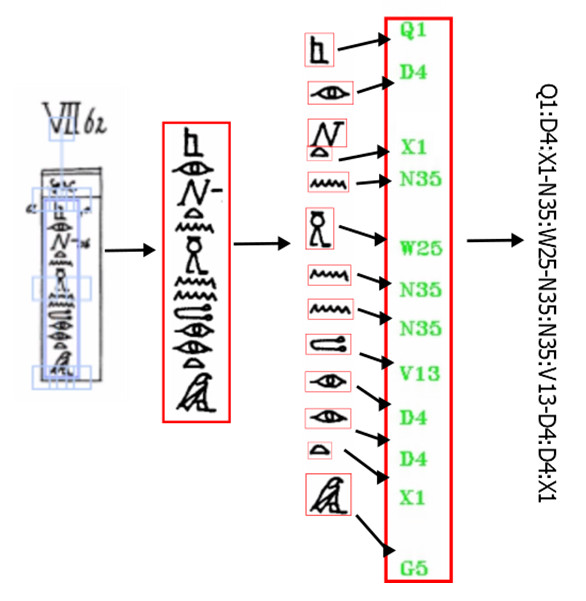} 
	\caption{Proposed transcription workflow using facsimiles as input.}\label{fig_5}
\end{figure}

\begin{figure}[!b]
	\centering
	
	\subfloat[G5]{%
		\includegraphics[width=0.10\linewidth]{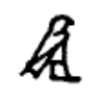}
		
	}
	\hspace{0.02\linewidth}
	\subfloat[D21]{%
		\includegraphics[width=0.10\linewidth]{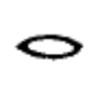}
		
	}
	\hspace{0.02\linewidth}
	\subfloat[Z1]{%
		\includegraphics[width=0.10\linewidth]{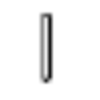}
		
	}
	\hspace{0.02\linewidth}
	\subfloat[W10]{%
		\includegraphics[width=0.10\linewidth]{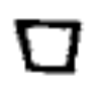}
		
	}	
	\vspace{2mm}
	
	\subfloat[G14]{%
		\includegraphics[width=0.10\linewidth]{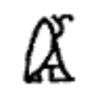}
		
	}
	\hspace{0.02\linewidth}
	\subfloat[Aa15]{%
		\includegraphics[width=0.10\linewidth]{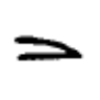}
		
	}
	\hspace{0.02\linewidth}
	\subfloat[U33]{%
		\includegraphics[width=0.10\linewidth]{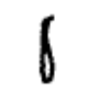}
		
	}
	\hspace{0.02\linewidth}
	\subfloat[Q3]{%
		\includegraphics[width=0.10\linewidth]{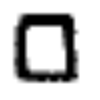}
		
	}	
	
	\caption{}
	\label{fig_11}
\end{figure}

After classification, the resulting Gardiner codes are assembled into transcription lines that respect the original column and token structure. The system incorporates a post-processing module that automatically builds a CSV file that records, for each entry, the support (coffin or pyramid), the token, the spell and the transcribed code. All this process is depicted in Fig. \ref{fig_5}. Furthermore, as an example, Table \ref{tab4} shows the transcription resulting from the ground-truth page in Fig. \ref{fig_4}, where each row corresponds to a column in the source image. In the proposed transcription, those codes that were wrongly identified are highlighted in red. Furthermore, it is worth noting that the sign \textit{\textbf{N}}, as it is not actually a hieroglyph but an abbreviation by the author in \cite{Buck61}, has also been discarded in the segmentation process.
As can be observed in Table \ref{tab4}, the results are consistent with the numerical figures obtained before. Furthermore, it is worth noting that, whether some hieroglyphs wrongly identified, the cause is likely a certain degree of likelihood between the correct hieroglyph and the misclassified one. For clarity's sake, Fig. \ref{fig_11} shows examples for the wrong classifications happening in the different columns showed in Table \ref{tab4}. As can be observed, most part of the errors are because geometrical similarities liked W10 and Q3 or macroclass similarities like G5 and G14 which both belongs to G macroclass. Usually these hieroglyphs are pretty similar, which makes it difficult to distinguish them for the proposal.

\begin{table}[htbp]
	\caption{Transcription results obtained with the previously proposed workflow for the ground-truth page shown in Fig. \ref{fig_4}. The hieroglyphs wrongly identified are highlighted in red. }
	\centering
	
	\includegraphics[width=0.70\linewidth]{figs/test_1_eg.PNG} 
	
	\vspace{0.3cm} 
	
	\resizebox{\textwidth}{!}{%
		\begin{tabular}{|c|c|c|}
			\hline
			\textbf{Column} &   \textbf{Proposed transcription}  &   \textbf{Ground-Truth} \\
			\hline
			(CT VII 62 a) &   Q1:D4:X1-N35:W25-N35:N35:V13-D4:D4:X1:\textcolor{red}{G14} & Q1:D4:X1-N35:W25-N35:N35:V13-D4:D4:X1:G5\\
			\hline
			(CT VII 62 b) & Q3:I10*X1-\textcolor{red}{N1}:\textcolor{red}{U33}*F34-I9 & Q3:I10*X1-T9D:Z1*F34-I9  \\
			\hline
			-Q3:I10*X1-N1:U33*F34-I9
			(CT VII 62 c) & N35:M37:\textcolor{red}{Aa15}-N35:V13:S29-N35:M17*D36-D58*\textcolor{red}{Q3}-N35:V13:S29-N35:X1:U15:G17 & N35:M37:D21-N35:V13:S29-N35:M17*D36-D58*W10-N35:V13:S29-N35:X1:U15:G17 \\
			\hline
			(CT VII 62 d) & N35:G5-Z11*Q1-D4:X1:N35 &  N35:G5-Z11*Q1-D4:X1:N35 \\
			\hline
			(CT VII 62 e) & Q1:D4:X1:N35:W25-N35:N35:V31-D4:D4:X1:G5  & Q1:D4:X1:N35:W25-N35:N35:V31-D4:D4:X1:G5 \\
			\hline
			(CT VII 62 f) & Q3:I10*X1-\textcolor{red}{T9}:F34*D50:I9 & Q3:I10*X1-T9D:F34*D50:I9 \\
			\hline
			(CT VII 62 g) & N35:M37:\textcolor{red}{Aa15}-N35:V31:S29-N35:M17-D36:D58*W10-N35:V31:S29:N35  & N35:M37:D21-N35:V31:S29-N35:M17-D36:D58*W10-N35:V31:S29:N35\\
			\hline
		\end{tabular}%
	}
	\label{tab4}
\end{table}

\subsection{Explainability in Deep Metric Learning}

\label{sec:explainability_tsne}

The next objetive is to obtain an interpretable representation of the proposed approximation. This representation will help us to supervised what the system have learn and the possible weak points in terms of errors or confusions. To achieve the interpretability of the proposed approximation, which is Deep-MML, we analyze the structure of its learned embedding space in 2D. Deep-MML maps each input sign image $I$ to an L2-normalized descriptor $\mathbf{z}=f(I)\in\mathbb{R}^{d}$, where classification is performed by comparing $\mathbf{z}$ against class template mean centroids using cosine similarity. While quantitative results summarize performance, visualizing the embedding distribution provides qualitative insight into: a) intra-class variability, b) inter-class separability, and c) systematic confusions between visually similar signs.

\begin{figure}
	\centering
	\includegraphics[width=1.\linewidth]{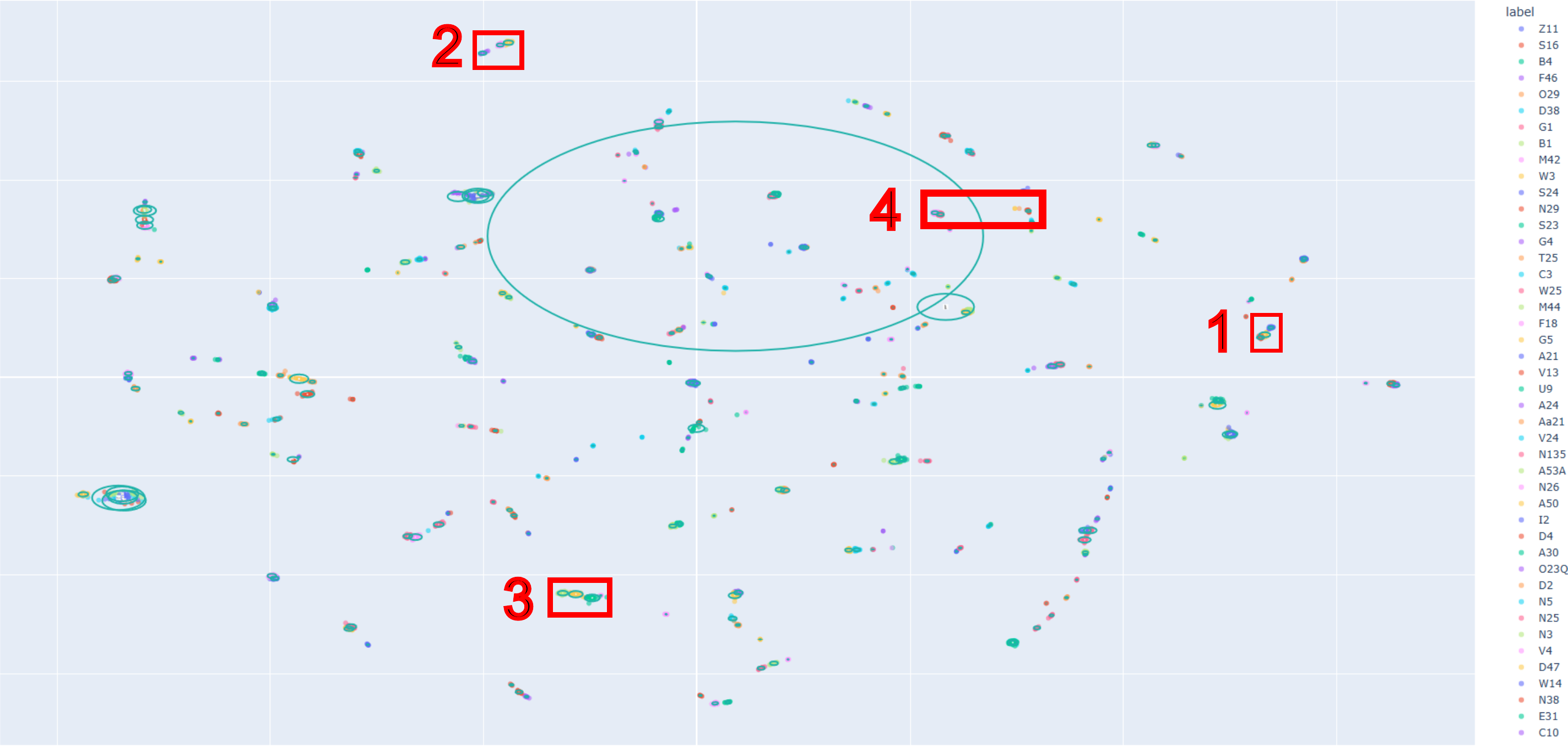} 
	\caption{T-SNE visualization of the latent space in 2D.}\label{fig:interpretability}
\end{figure}

For this purpose, we compute embeddings for samples from the evaluation set and project them into two dimensions using t-distributed Stochastic Neighbor Embedding (t-SNE). The resulting 2D maps are shown in Figure \ref{fig:interpretability} using Plotly \cite{kruchten2025plotly}, where each circle represents one hieroglyph instance, colors denote classes and each ellipse and centroid models the class centroid and distribution along the latent space. Clusters or aggrupations with compact structure indicate low intra-class variability in the learned representation, whereas overlapping neighborhoods reveal signs that are close in the embedding space and therefore prone to confusion.

We have marked 4 different zones in Figure \ref{fig:interpretability}, that allow us to use this representation to model the errors resulted from \ref{fig_11}. If we study case by case the representation, we can see clearly the relation between the representation and the errors.
\begin{itemize}
	\item 1: The confusion between G14 and G5 is modeled by the representation, we can see that these 2 classes are overlapped in the representation space as Figure \ref{fig_12} show, being the G14 the bottom left one and the G5 the center one.
	\begin{figure}
		\centering
		\includegraphics[width=1.\linewidth]{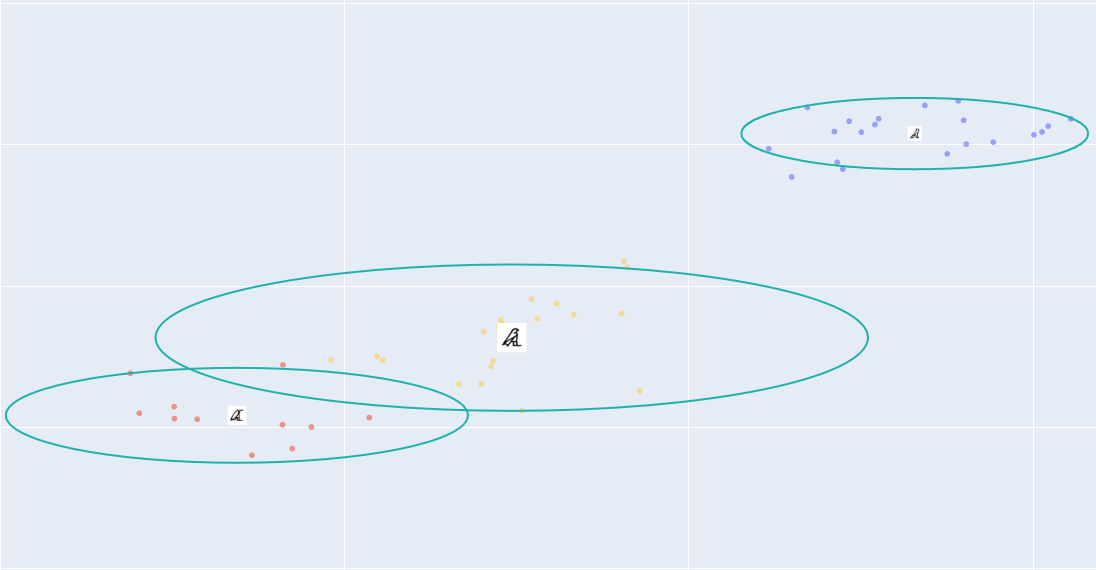} 
		\caption{Confusion case G14-G5.}\label{fig_12}
	\end{figure}
	\item 2: The confusion between G14 and G5 is modeled by the representation, we can see that these 2 classes are not overlapped in the representation space but very near increasing the probable confusion between them as Figure \ref{fig_13} show, being the D21 the bottom left one and the Aa15 the upper right center one.
	\begin{figure}
		\centering
		\includegraphics[width=1.\linewidth]{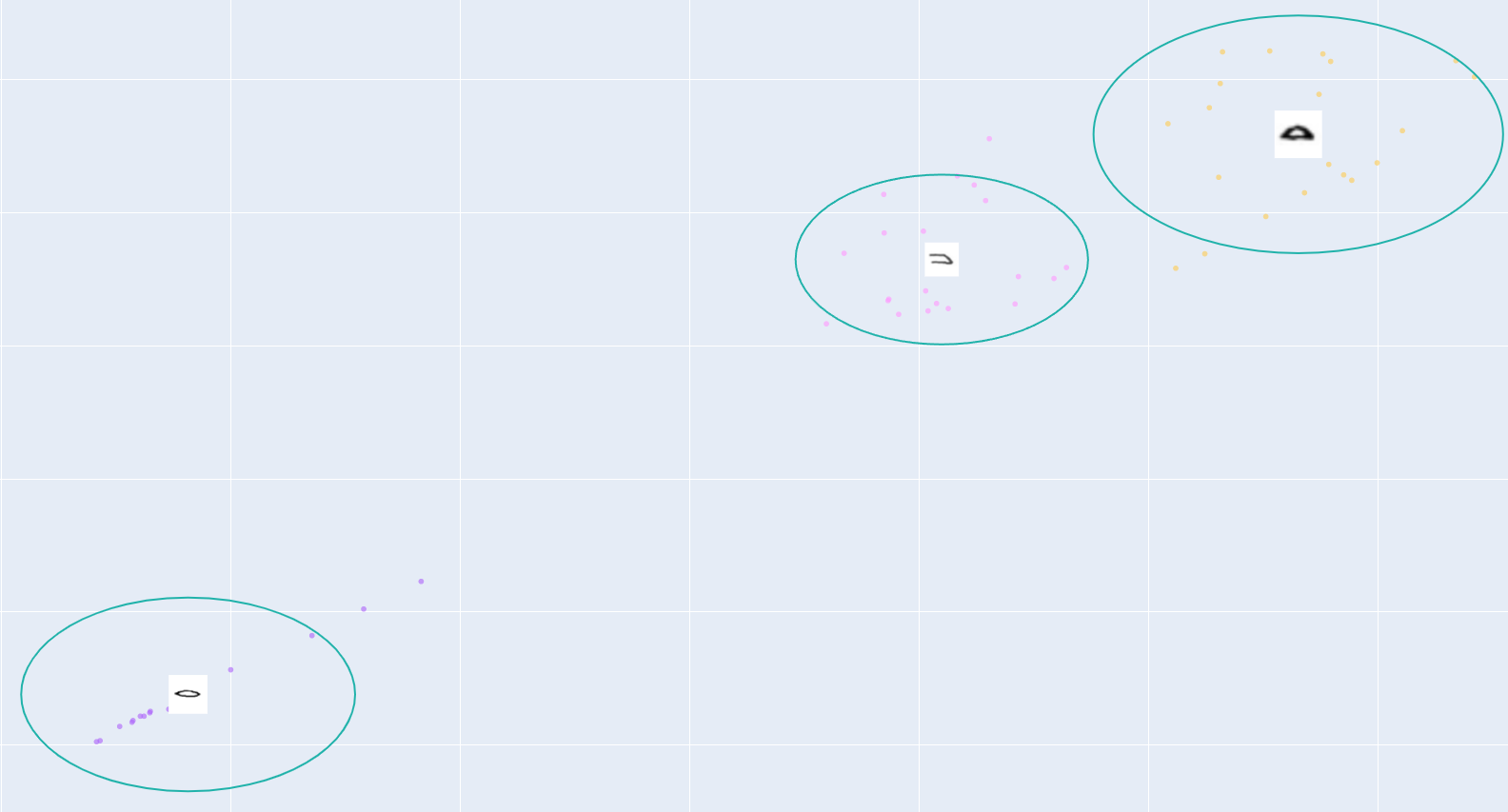} 
		\caption{Confusion case D21-Aa15.}\label{fig_13}
	\end{figure}
	\item 3: The confusion between Z1 and U33 is modeled by the representation, we can see that these 2 classes are not overlapped in the representation space but very near  as Figure \ref{fig_14} show, being the U33 the left one and the Z1 the right one. This case is very complicated compared with the others because these signs don't have a lot of information, and in terms of shape they are very similar, taking in account that the image of facsimiles usually have poor resolution for each symbol.
	\begin{figure}
		\centering
		\includegraphics[width=1.\linewidth]{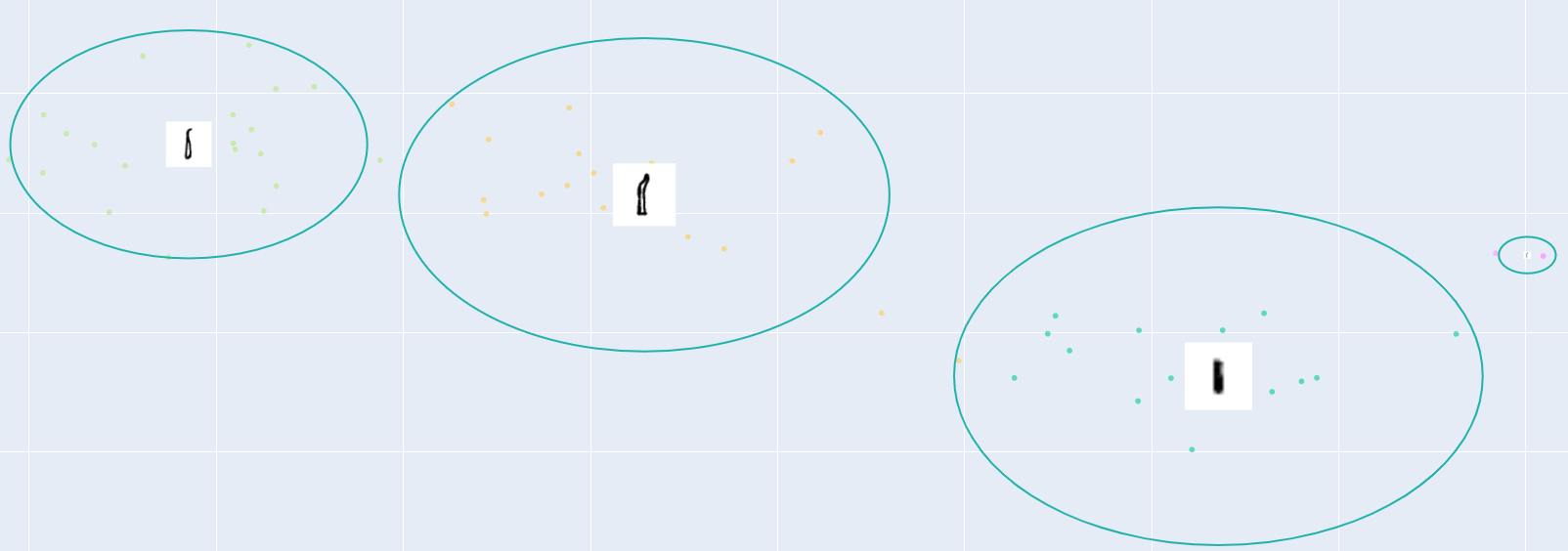} 
		\caption{Confusion case Z1-U33.}\label{fig_14}
	\end{figure}
	
	\item 4: The confusion between W10 and Q3 is an strange case, we can see that in the mapping the clases are not very near, so is not easy compared with other cases to extrapolate in terms of the latent space that is a very probable error. If we visualize the symbols we can see easily that the problem comes from the shape of the two hieroglyphs that is very similar to squares if the are not very well drawed. Figure \ref{fig_15} show the relation between these two, being Q3 the right one and W10 the bottom left one.
	\begin{figure}
		\centering
		\includegraphics[width=1.\linewidth]{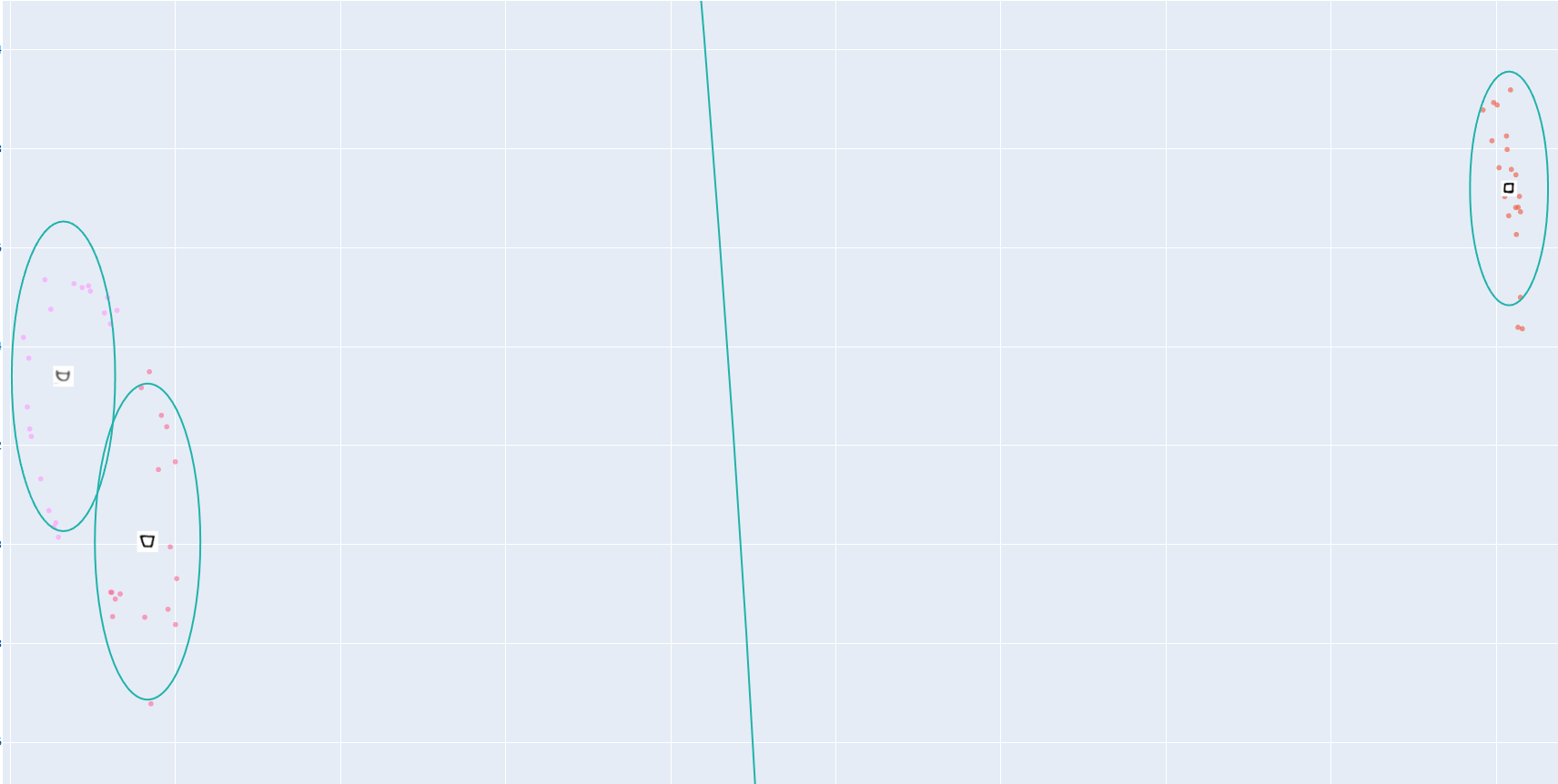} 
		\caption{Confusion case Q3-W10.}\label{fig_15}
	\end{figure}
\end{itemize}

\section{Conclusions}
\label{sec7}

This study has dealt with the definition of different proposals within the field of machine learning, capable of processing images with texts from coffins and/or pyramids, in order to identity the hieroglyphs, transcript them to Gardiner’s codes, and store them for further analysis. For that purposes, three approaches have been considered: a classical machine learning one, called as Traditional Machine Learning (\method{Trad-ML}) and based on SVM with pre-selected input features; a classical deep learning approach, named as End-to-End Deep CNN Classifier (\method{CNN-End2End}) and which uses feature extractors and convolutional classifiers; and, finally, our proposed Deep Metric Learning model (\method{Deep-MML}), using a trained feature extractor to differentiate hieroglyphs, not necessarily trained, in a multidimensional space. Among all, the latter proposal has proven a remarkable capability for semi-automatic classification of Egyptian hieroglyphs, outperforming traditional approaches, such as SVM and CNN classifiers, in scenarios with a strong imbalance and/or a lack of class samples. With an overall accuracy of $98.8$\% on the test set and $88.9$\% on some reference pages not seen during training, the metric learning method guarantees higher stability against undersampled classes and facilitates the incorporation of new signs without the need to retrain the whole model. Accordingly, Deep-MML is used as the default recognition engine in the released OCR-PT-CT application. Finally, it is worth noting that all this workflow has been integrated in a web tool, which allows csv files to be generated with all the information involved in the process.

As lines of future work, it is proposed to explore these cases on images taken with cameras on walls full of hieroglyphs by applying a style change to facsimile beforehand. Similarly, there is significant potential in investigating more robust segmentation architectures, e.g., based on zero-shot segmentation networks such as SAM2\cite{ravi2025sam}, to improve the separation of hieroglyphs on low quality supports. Finally, the extension of the system to other sets of ancient texts and the direct fusion with automatic transliteration modules will consolidate a comprehensive environment for the digital study of Egyptian languages, opening new avenues for interdisciplinary collaboration.

\section{Acknowledgments}
This work was supported in part by the Spanish Ministry of Science, Innovation and Universities MICIU/AEI/10.13039/501100011033 (project METAMORPH, ref. PID2023-151295OB-I00), the University of Alcala (project AMPITET, ref. PIUAH24/IA-016), the Spanish Ministry for Research through the TM-CT project under grant CNS2023-144010 and the Castilla-La Mancha Region through the T3D project under grant SBPLY/23/180225/000093. The authors would like to thank University of Chicago for their colaboration with Carlos Gracia and Álvaro Rojo for his participation in the processing of the corpus data used by the system.

\bibliographystyle{ieee}
\bibliography{biblio}

\end{document}